\documentclass[natbib,twocolumn,final]{svjour3}

\usepackage{url}
\usepackage{doi}
\usepackage{amssymb}
\usepackage{amsmath}
\usepackage{graphicx}

\begin{document}\sloppy

\def\makeheadbox{\relax}

\title{JIGSAW-GEO (1.0): locally orthogonal staggered unstructured grid generation for general circulation modelling on the sphere\thanks{A short version of this paper appears in the proceedings of the 24th International Meshing Roundtable \citep{engwirda2015multi}.}}

\author{Darren Engwirda}

\institute{D.~Engwirda \at 
Department of Earth, Atmospheric and Planetary Sciences, Room 54-1517, Massachusetts Institute of Technology, 77 Massachusetts Avenue, Cambridge, MA 02139-4307, USA \\
Corresponding author. Tel.: +1-212-678-5521. \\
\email{engwirda@mit.edu; de2363@columbia.edu}
\and
NASA Goddard Institute for Space Studies, 2880 Broadway, New York, NY 10025 USA}

\titlerunning{Unstructured grid generation for general circulation modelling}

\date{Published: 6 June, 2017, \url{https://doi.org/10.5194/gmd-10-2117-2017}}

\maketitle

\begin{abstract}
An algorithm for the generation of non-uniform, locally-orthogonal staggered unstructured spheroidal grids is described. This technique is designed to generate very high-quality staggered Voronoi/Delaunay meshes appropriate for general circulation modelling on the sphere, including applications to atmospheric simulation, ocean-modelling and numerical weather prediction. Using a recently developed Frontal-Delaunay refinement technique, a method for the construction of high-quality unstructured spheroidal Delaunay triangulations is introduced. A locally-orthogonal polygonal grid, derived from the associated Voronoi diagram, is computed as the staggered dual. It is shown that use of the Frontal-Delaunay refinement technique allows for the generation of very high-quality unstructured triangulations, satisfying a-priori bounds on element size and shape. Grid-quality is further improved through the application of hill-climbing type optimisation techniques. Overall, the algorithm is shown to produce grids with very high element quality and smooth grading characteristics, while imposing relatively low computational expense. A selection of uniform and non-uniform spheroidal grids appropriate for high-resolution, multi-scale general circulation modelling are presented. These grids are shown to satisfy the geometric constraints associated with contemporary unstructured C-grid type finite-volume models, including the Model for Prediction Across Scales (MPAS-O). The use of user-defined mesh-spacing functions to generate smoothly graded, non-uniform grids for multi-resolution type studies is discussed in detail.

\end{abstract}


\section{Introduction}
\label{section_introduction}

The development of atmospheric and oceanic general circulation models based on \textit{unstructured} numerical discretisation schemes is an emerging area of research. This trend necessitates the development of unstructured grid-generation algorithms designed to produce very high-resolution, guaranteed-quality unstructured triangular and polygonal meshes that satisfy non-uniform mesh-spacing distributions and embedded geometrical constraints. This study investigates the applicability of a recently developed surface meshing algorithm \citep{engwirda2016157,engwirda2016conforming} based on restricted Frontal-Delaunay refinement and hill-climbing type optimisation for this task.

\subsection{Semi-structured grids}

\begin{figure*}
  \centering
  
  \includegraphics[width=.915\textwidth]{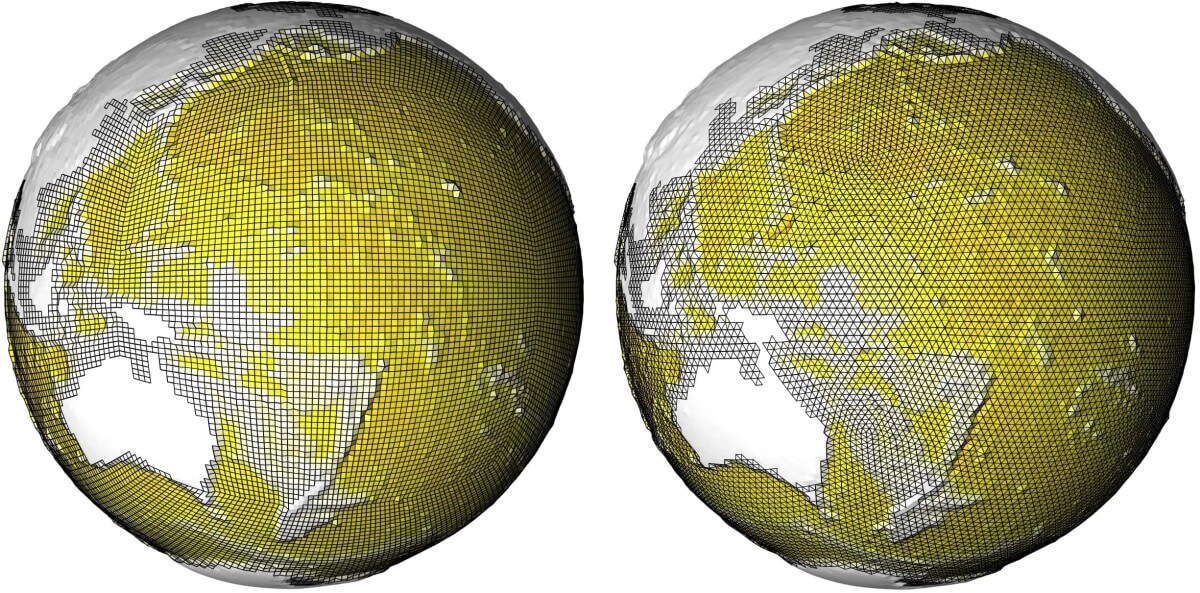}
  
  \caption{Conventional semi-structured meshes for the sphere, showing a cubed-sphere grid (left), and an icosahedral grid (right). Both grids were generated using equivalent target mean edge lengths, and are coloured according to mean topographic height at grid-cell centres. Topography is drawn using an exaggerated scale, with elevation from the reference geoid amplified by a factor of 20 in both cases.}
  \label{figure_structured_grids}
\end{figure*}

While simple structured grid types for the sphere can be obtained by assembling a uniform discretisation in spherical coordinates, the resulting \textit{lat-lon} grid is often inappropriate for numerical simulation, due to the presence of strong \textit{grid-singularities} at the two poles. Such features manifest as local distortions in grid-quality, consisting of regions of highly distorted quadrilateral grid-cells. These low-quality elements can lead to a number of undesirable numerical effects --- imposing restrictions on model time-step and stability, and compromising local spatial accuracy. As a result, a majority of current generation general circulation models are instead based on \textit{semi-structured} quadrilateral discretisation schemes, including the \textit{cubed-sphere} \citep{adcroft2004implementation,marshall1997finite,putman2007finite} and \textit{tri-polar} configurations \citep{murray1996explicit,madec2015nemo,bleck2002oceanic}. 

In the cubed-sphere framework, the spherical surface is decomposed into a cube-like topology, with each of the six quadrilateral faces discretised as a structured curvilinear grid. In such an arrangement, the two strong grid-singularities of the lat-lon configuration are replaced by eight weak discontinuities at the cube corners, leading to significant improvements in numerical performance. \citet{putman2007finite} present detailed discussions of techniques for the generation and optimisation of cube-sphere type grids. A regular \textit{gnomonic-type} cubed-sphere grid is illustrated in Figure~\ref{figure_structured_grids}a. 

In the tri-polar grid, the present-day continental configuration is exploited to \textit{bury} the singularities associated with a three-way polar decomposition of the sphere outside of the ocean mask. The resulting \textit{numerically-active} subset of the grid is well-conditioned as a result. While such configurations are a popular choice for models designed for present-day Earth-centric ocean studies, the generality of these methods is clearly limited. In this study, we instead pursue the development of more general-purpose techniques, applicable to both oceanic and atmospheric modelling for general planetary and paleo-Earth environments. 

In addition to the standard cubed-sphere and tri-polar configurations, a second class of semi-structured spherical grid can be constructed through \textit{icosahedral-type} decompositions \citep{heikes1995numerical,randall2002climate}. In such cases, the primary grid is defined as a regular spherical triangulation, obtained through recursive bisection of the icosahedron. The associated staggered polygonal \textit{dual} grid, consisting of hexagonal and pentagonal cells, is often used as a basis for finite-volume type numerical schemes. This \textit{geometric-duality} is an example of the locally-orthogonal Delaunay/Voronoi type grid staggering that forms the basis of this paper. Icosahedral-type grids provide a near-perfect tessellation of the sphere --- free of topological discontinuities and/or geometrical irregularity. Such methods are applicable to both atmospheric and oceanic type simulations. A regular icosahedral-class grid is illustrated in Figure~\ref{figure_structured_grids}b. 

\subsection{Unstructured grids}

While the semi-structured grids described previously each provide effective frameworks for uniform resolution global simulation, the development of \textit{multi-resolution} modelling environments requires alternative techniques. A range of new general circulation models, including the Finite Element Sea Ice-Ocean Model (FESOM) \citep{gmd-7-663-2014}, the Finite Volume Community Ocean Model (FVCOM) \citep{chen2003unstructured,chen2007finite,lai2010nonhydrostatic}, the Stanford Unstructured Non-hydro\-static Terrain-following Adaptive Navier-Stokes Simulator (SUNTANS) \citep{fringer2006unstructured,vitousek2014nonhydrostatic}, and the Second-generation Louvain-la-Neuve Ice-ocean Model (SLIM) \citep{bernard2007high,comblen2009finite} are based on semi-structured triangular grids, with the horizontal directions discretised according to an unstructured spherical triangulation, and the vertical direction represented as a stack of locally structured layers. The Model for Predication Across Scales (MPAS) \citep{skamarock2012multiscale,ringler2013multi,ringler2008multiresolution} adopts a similar arrangement, except that a \textit{locally-orthogonal} unstructured discretisation is adopted, consisting of both a Spherical Voronoi Tessellation (SVT) and its dual Delaunay triangulation. The use of fully unstructured representations, based on general tetrahedral and/or polyhedral grids, are also under investigation in the Fluidity framework \citep{ford2004nonhydrostaticA,ford2004nonhydrostaticB,pain2005three,piggott2008new}. Such models all impose different requirements on the \textit{quality} of the underlying unstructured grids, with some models, including FESOM, SLIM and Fluidity, offering additionally flexibility. In all cases though, the performance of the numerical simulation can be expected to improve with increased grid-quality --- encouraging the search for optimised grid-generation algorithms. Further discussion of grid-quality constraints for general circulation modelling is presented in Section~\ref{section_mpas_grids}. 

Existing approaches for unstructured grid-generation on the sphere have focused on a number of techniques, including: (i) the use of iterative, optimisation-type algorithms designed to construct Spherical Centroidal Voronoi Tessellations (SCVT's) \citep{jacobsen2013parallel}, and (ii) the adaptation of anisotropic two-dimensional meshing techniques \citep{lambrechts2008multiscale} that build grids in associated parametric spaces. 

The MPI-SCVT algorithm \citep{jacobsen2013parallel} is a massively parallel implementation of iterative Lloyd-type smoothing \citep{du1999centroidal} for the construction of SCVT's for use in the MPAS framework. In this approach, a set of vertices are distributed over the spherical surface and iteratively \textit{smoothed} until a high-quality Voronoi tessellation is obtained. Specifically, each iteration repositions vertices to the centroids of their associated Voronoi cells and updates the topology of the underlying spherical Delaunay triangulation. While such an approach typically leads to the generation of high-quality \textit{centroidal} Voronoi tessellations, the algorithm does not provide theoretical guarantees on minimum element quality, and often requires significant computational effort to achieve convergence. Additionally, current implementations of the MPI-SCVT algorithm do not provide a mechanism to constrain the grid to embedded features, such as coastal boundaries.

\begin{figure*}
  \centering
  \includegraphics[width=.800\textwidth]{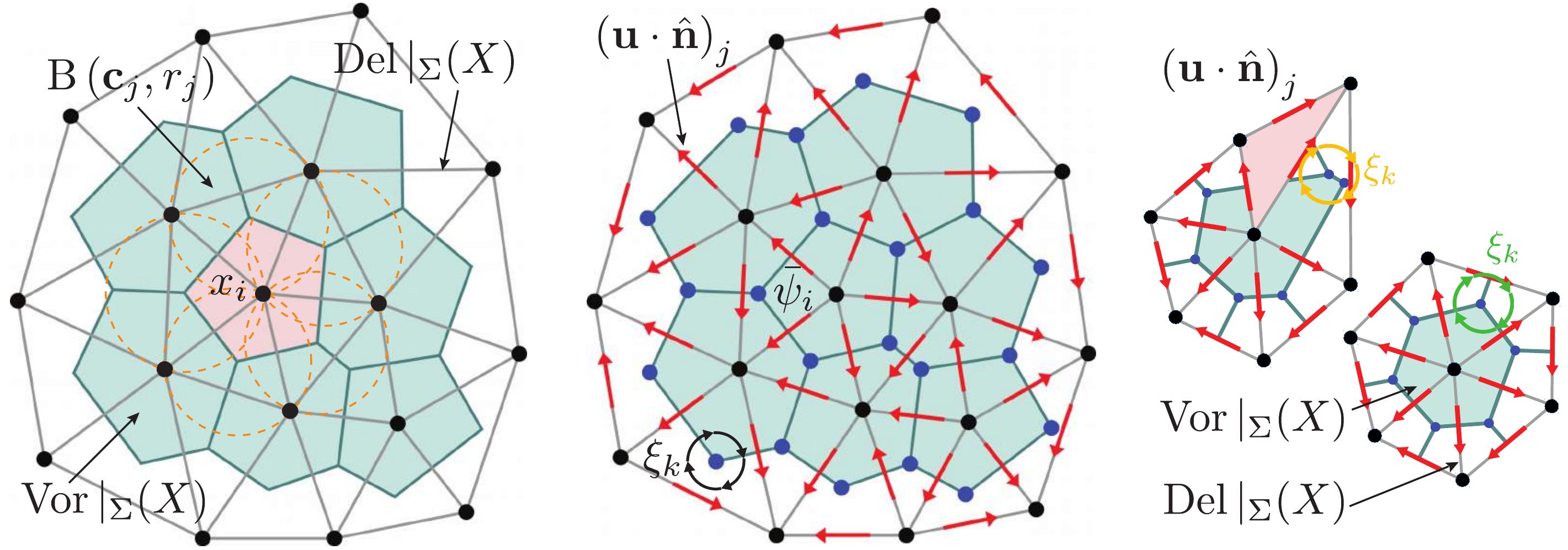}
  
  \caption{An anatomy of staggered unstructured grids for general circulation modelling, illustrating: (a) locally-orthogonal Voronoi/Delaunay grid staggering, (b) an unstructured C-grid type numerical formulation, and (c) a comparison of \textit{well-centred} and \textit{poorly-staggered} configurations. In (a), Voronoi polygons are formed by joining the centres of circumscribing balls $\operatorname{B}(\mathbf{c}_{i},r_{i})$ associated with adjacent Delaunay triangles. In (b), the C-grid scheme consists of conservative \textit{cell-centred} tracer quantities $\bar{\psi}_{i}$, \textit{edge-centred} normal velocity components $(\mathbf{u}\cdot\hat{\mathbf{n}})_{j}$, and auxiliary \textit{vertex-centred} vorticity variables $\xi_{k}$. In (c), the \textit{well-centred} (lower) and \textit{poorly-staggered} (upper) configurations differ in the associativity of Voronoi vertices. In the well-centred configuration, Voronoi vertices are \textit{interior} to their parent triangles. Reconstruction of the vertex-centred vorticity distribution can thus be achieved by integrating over the Delaunay triangles. In the poorly-staggered configuration, a single Voronoi vertex is exterior to its parent triangle (shaded), leading to a breakdown in the reconstruction. Note that adjacent Voronoi/Delaunay edges do not intersect in the poorly-staggered configuration.
  }
  \label{figure_voronoi_construction}
\end{figure*}

\citet{lambrechts2008multiscale} present an unstructured spherical triangulation framework using the general-purpose grid-generation package Gmsh \citep{geuzaine2009gmsh}. In this work, unstructured spherical triangulations are generated for the world ocean using a parametric meshing approach. Specifically, a triangulation of the spherical surface is generated by \textit{mapping} the full domain (including coastlines) on to an associated two-dimensional parametric space via stereographic projection. Importantly, as a result of the projection, the grids constructed in parametric space must be highly anisotropic, such that a well-shaped, isotropic triangulation is induced on the sphere. A range of existing two-dimensional anisotropic meshing algorithms are investigated, including Delaunay-refinement, advancing-front, and adaptation-type approaches. In particular, the algorithm is designed to ensure a faithful representation of complex coastal boundary conditions. While a detailed model of such constraints is often neglected in global simulations, resolution of these features is a key factor for regional and coastal models. Several additional algorithms support the generation of unstructured grids for such two-dimensional domains, including the ADmesh package \citep{conroy2012admesh} and the Stomel library \citep{holleman2013numerical}.

The current study describes a new algorithm for the generation of guaranteed-quality spheroidal Delaunay triangulations and Voronoi tessellations --- appropriate for a range of unstructured  general circulation models. In this work, meshes are generated on the spheroidal surface directly, without need for local parameterisation or projection. Such an approach will be shown to exhibit significant flexibility --- immune to issues of coordinate singularity and/or continental configuration. The applicability of this approach to grid-generation for imperfect spheres, including oblate spheroids and general ellipsoids is also discussed. Significant effort is invested to develop techniques designed to produce very high-quality multi-resolution grids appropriate for contemporary unstructured C-grid models, including the Model for Prediction Across Scales (MPAS). 

The paper is organised as follows: an overview of grid-generation for general circulation modelling is presented in Section~\ref{section_mpas_grids}, outlining various constraints and requirements on minimum grid-quality. A description of the Frontal-Delaunay refinement and hill-climbing type optimisation algorithms is given in Sections~\ref{section_refinement_algorithm} and \ref{section_optimisation_algorithm}. A set of uniform and non-uniform grids appropriate for high-resolution, multi-scale general circulation modelling are presented Section~\ref{section_results}, alongside an analysis of computational performance and optimality. Avenues for future work are outlined in Section~\ref{section_conclusion}.

\section{Grid-generation for general-circulation modelling}
\label{section_mpas_grids}

Numerical formulations for atmospheric and/or oceanic general circulation modelling are typically based on a \textit{staggered} grid configuration, with quantities such as fluid pressure, geopotential height, and density discretised using a primary control-volume, and the fluid velocity field and vorticity distribution represented at secondary, spatially distinct grid-points. Various structured staggered grid types are described by the well-known Arakawa schemes \citep{arakawa1977computational}.

The development of models based on unstructured grid types is an emerging area of research, and, as a result, a variety of numerical formulations are currently under investigation. In this study, the use of \textit{locally-orthogonal} grids appropriate for staggered unstructured C-grid schemes is pursued, as these methods represent a logical extension of conventional structured Arakawa-type techniques to the unstructured setting. Such formulations require that the grids satisfy a \textit{local-orthogonality} constraint, with adjacent grid-cell edges in the primary and secondary control-volumes required to be mutually perpendicular. In the unstructured setting, it is known that the Delaunay triangulation and the Voronoi tessellation constitute a locally-orthogonal staggered dual, leading to a natural framework for the construction of such meshes.

Given a Delaunay triangulation $\operatorname{Del}|_{\Sigma}(X)$ of the surface $\Sigma$, the associated Voronoi diagram $\operatorname{Vor}|_{\Sigma}(X)$ can be constructed by exploiting this underlying \textit{geometric-duality}. Each edge in the Delaunay triangulation is associated with a \textit{dual-edge} in the Voronoi diagram, spanning between the centres of the circumscribing-balls $\operatorname{B}(\mathbf{c}_{j},r_{j})$ associated with the two adjacent Delaunay triangles. The full Voronoi diagram is assembled from these edges --- consisting of a set of (convex) polygonal grid-cells centred on each vertex in the underlying surface triangulation. See Figure~\ref{figure_voronoi_construction}a for an example of this unstructured Voronoi/Delaunay type grid staggering.

In Figure~\ref{figure_voronoi_construction}b, an unstructured variant of the widely-used Arakawa C-grid scheme is described, with fluid pressure, geopotential height and density degrees-of-freedom placed within the primary Voronoi control-volumes, and a set of orthogonal velocity vectors positioned along Delaunay edges. Additional vorticity degrees-of-freedom are located at the vertices of the Voronoi grid-cells. Such an arrangement facilitates the construction of a standard conservative finite-volume type scheme for the transport of cell-centred fluid properties, and a \textit{mimetic} class \citep{lipnikov2014mimetic,bochev2006principles} finite-difference formulation for the evolution of velocity components. Overall, this scheme is known to posses a variety of desirable conservation properties, conserving mass, potential vorticity and enstrophy, and preserving geostrophic balance \citep{ringler2010unified}. This unstructured C-grid scheme is currently employed in the Model for Prediction Across Scales (MPAS) for both atmospheric and oceanic modelling \citep{skamarock2012multiscale,ringler2013multi,ringler2008multiresolution}.

While numerically elegant, such schemes are not applicable to general unstructured Voronoi/Delaunay grids --- requiring that a number of auxiliary geometrical constraints also be satisfied. Specifically, such schemes necessitate the use of grids that are not only locally-orthogonal, but are also \textit{well-centred} and mutually \textit{centroidal}. These additional conditions are constraints on the geometry of the underlying Delaunay triangles and Voronoi polygons.

A \textit{well-centred} Delaunay grid is one in which all dual Voronoi vertices lie within the interior of their associated Delaunay triangles. Such a condition guarantees that adjacent Delaunay and Voronoi edges intersect --- inducing a dual structure that is \textit{nicely} staggered. In the context of the unstructured C-grid scheme described previously, this configuration guarantees that a consistent stencil exists for reconstruction of the discrete vorticity variable. In practice, the generation of well-centred grids is known to be a particularly onerous task \citep{vanderzee2008well,vanderzee2010well}, requiring that the triangulation consist of \textit{all-acute} elements. Degradation of the vorticity reconstruction for poorly-staggered grids is described in Figure~\ref{figure_voronoi_construction}c.

Delaunay grids are \textit{centroidal} when their primary and secondary vertices lie at the centres-of-mass of their associated dual grid-cells, with the vertices of the Voronoi polygons lying at the centroids of the Delaunay triangles and visa-versa. Such a condition is effectively an implicit constraint on the \textit{regularity} of the grid, with strongly centroidal meshes typically associated with improved grid-cell shapes. In the limit, a perfectly centroidal and well-centred tessellation can be assembled from equilateral triangles and regular hexagons. Centroidal grids typically lead to very high-quality numerical discretisations, with grids containing near-perfect element configurations achieving \textit{optimal} convergence rates. The unstructured C-grid scheme described previously is known to achieve second-order accurate convergence when applied to centroidal Voronoi grids \citep{ringler2010unified}.

The generation of \textit{locally-orthogonal}, \textit{well-centred} and \textit{centroidal} unstructured grids is a difficult task, imposing a heavy-burden on the underlying mesh generation algorithm. While a number of techniques for unstructured grid-generation currently exist, as per Section~\ref{section_introduction}, I am not aware of any that are successful in generating the very high-quality grids required by unstructured C-grid type general circulation models. As a result, throughout the remainder of this paper, the development of new methods for the generation of such tessellations is pursued in detail.

\section{A restricted Frontal-Delaunay refinement algorithm}
\label{section_refinement_algorithm}

The task is to generate very high-resolution, guaranteed-quality unstructured Delaunay triangulations for planetary atmospheres and/or oceans. These grids will form a baseline for the hill-climbing mesh-optimisation techniques presented in subsequent sections. Such grids are required to satisfy a number of constraints, including: bounds on minimum element quality and adherence to user-defined mesh-spacing distributions. In this work, the applicability of a recently developed Frontal-Delaunay surface meshing algorithm \citep{engwirda2016157,engwirda2016conforming} is investigated for this task.

An unstructured Delaunay triangulation of the reference spheroid associated with a general planetary geometry is sought. In a general form, this reference surface can be expressed as an axis-aligned triaxial ellipsoid
\begin{equation}
\label{equation_ellipsoid}
\qquad \sum_{i=1}^{3}\, \left(\frac{x_{i}}{r_{i}}\right)^{2} = 1 \,,
\end{equation}

where the $x_{i}$'s are the Cartesian coordinates in a locally aligned coordinate system, and the scalars $r_{i} > 0$ are its principal radii. Such a definition can be used to represent ellipsoidal surfaces in general position, based on the application of additional rigid-body translations and rotations. Note that while grid-generation for global climate modelling is often restricted to spherical surfaces (e.g.~$r_{1,2,3}=6371\mathrm{km}$) this formulation facilitates mesh-generation for general spheroidal and ellipsoidal domains.

\subsection{Preliminaries}

Before describing the Frontal-Delaunay refinement algorithm in full, several key properties of the underlying Delaunay triangulation and Voronoi tessellation are briefly reviewed. A full account of these structures is not presented here, instead, the reader is referred to the detailed theoretical exposition presented in, for example, \citet{ChengDeyShewchuk}.

The Delaunay triangulation $\operatorname{Del}(X)$ associated with a set of points $X\in\mathbb{R}^{d}$ is characterised by the so-called \textit{empty-circle} criterion --- requiring that the set of circumscribing spheres $\operatorname{B}(\mathbf{c}_{i},r_i)$ associated with each Delaunay triangle $\tau_{i}\in\operatorname{Del}(X)$ be \textit{empty} of all points other than its own vertices. It is well known that for tessellations restricted to two-dimensional manifolds, the Delaunay triangulation leads to a maximisation of the minimum enclosed angle in the grid \citep{ChengDeyShewchuk}. Such behaviour is clearly beneficial when seeking to construct high quality triangular meshes. 

The Voronoi tessellation $\operatorname{Vor}(X)$ is the so-called \textit{geometric-dual} associated with the Delaunay triangulation, consisting of a set of convex polygonal cells formed by connecting the centres of adjacent circumscribing balls --- the so-called element \textit{circumcentres} $\mathbf{c}_{i}$'s. The Voronoi tessellation represents a \textit{closest-point map} for the points in $X$, with each Voronoi cell $v_{c}\in\operatorname{Vor}(X)$ defining the convex region adjacent to a given vertex $\mathbf{x}_{i}\in X$ for which $\mathbf{x}_{i}$ is the nearest point. Importantly, the Voronoi/Delaunay grid staggering, defines a \textit{locally-orthogonal} arrangement, in which grid-cell edges in the Voronoi tessellation are orthogonal to adjacent edges in the underlying Delaunay triangulation.

\subsection{Restricted Delaunay triangulation}

\begin{figure}
  \centering
  \includegraphics[width=.450\textwidth]{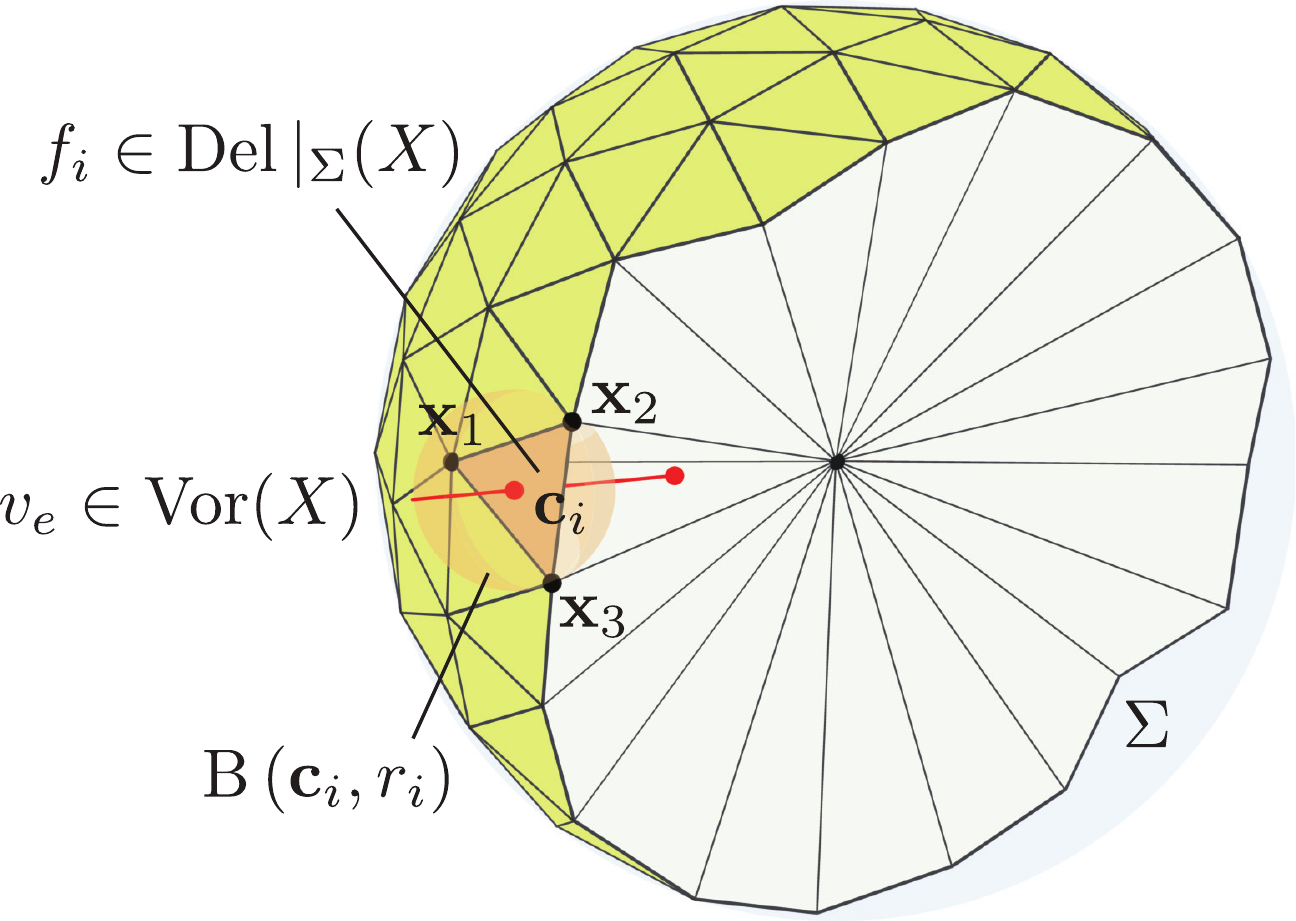}
  
  \caption{Illustration of the geometrical predicates used to define the restricted Delaunay surface triangules $f_{i}\in\operatorname{Del}|_{\Sigma}(X)$, showing details of the intersecting Voronoi edge $v_{e}\in\operatorname{Vor}(X)$ associated with the surface triangle $f_{i}\in\operatorname{Del}|_{\Sigma}(X)$ and its associated surface-ball $\operatorname{B}(\mathbf{c}_{i},r_{i})$. Note that the full three-dimensional Delaunay tessellation $\operatorname{Del}(X)$ is a tetrahedral complex that fills the interior of the spheroid.}
  \label{figure_restricted_faces}
\end{figure}

In this study, grid-generation is carried out directly on the spheroidal geometry by making use of so-called \textit{restricted} Delaunay mesh generation techniques. Specifically, given a reference surface $\Sigma$, grid-generation proceeds to discretise the surface into a mesh of triangles. In the restricted Delaunay framework, a full-dimensional Delaunay tessellation $\operatorname{Del}(X)$ (i.e.~a tetrahedral tessellation) is maintained, with the surface triangulation represented as a subset of the tetrahedral faces. The \textit{restricted} Delaunay surface triangulation $\operatorname{Del}|_{\Sigma}(X)$ is said to be \textit{embedded} in $\operatorname{Del}(X)$ as a result. Use of this fully three-dimensional approach elides any reliance on local parametric projections. 
\begin{definition}[restricted Delaunay tessellation]
\label{definition_restricted_tessellation}
Let $\Sigma$ be a smooth surface embedded in $\mathbb{R}^3$. Let $\operatorname{Del}(X)$ be a full-dimensional Delaunay tetrahedralisation of a point-wise sample $X\subseteq\Sigma$ and $\operatorname{Vor}(X)$ be the associated Voronoi tessellation. The \textit{restricted Delaunay surface triangulation} $\operatorname{Del}|_{\Sigma}(X)$ is a sub-complex of $\operatorname{Del}(X)$ including any triangle $f_{i}\in\operatorname{Del}(X)$ associated with an \textit{intersecting} Voronoi edge $\mathbf{v}_{e}\in\operatorname{Vor}(X)$ such that $\mathbf{v}_{e}\cap\Sigma\neq\emptyset$.
\end{definition} 
The development of restricted Delaunay techniques for general mesh-generation applications has been the subject of previous research, and a detailed discussion of such concepts is presented elsewhere. The reader is referred to the original work of \citet{Edelsbrunner97Restricted} or the detailed reviews presented in \citet{ChengDeyShewchuk} for additional details and mathematical background. 

In the context of this work, it is sufficient to note that the restricted Delaunay framework provides a convenient mechanism for the construction of grids on the spheroidal surface $\Sigma$, without requiring the introduction of parametric mappings or anisotropic metrics. Implementation of the restricted Delaunay scheme requires the definition of a single \textit{geometric-predicate} --- a local comparison used to determine which faces of the underlying tetrahedral complex lie on the spheroidal surface. Specifically, intersections between the edges of the full-dimensional Voronoi complex and the underlying surface $\Sigma$ are computed. Triangles associated with non-empty intersections $\operatorname{Vor}|_{f}(X)\,\cap\,\neq\emptyset$ form part of the surface mesh. In this study, these comparisons are computed analytically, following standard spheroidal trigonometric manipulations, as detailed in Appendix~\ref{appendix_predicates}. See Figure~\ref{figure_restricted_faces} for a detailed schematic.  

\subsection{Mesh-spacing functions}

The local \textit{density} of the mesh can be controlled via a user-specified mesh-spacing function $\bar{h}(\mathbf{x}) : \mathbb{R}^{3} \rightarrow \mathbb{R}^{+}$, where $\bar{h}(\mathbf{x})$ defines the \textit{target} edge-lengths at all points on the surface $\Sigma$. In this work, mesh-spacing functions are specified as a discrete set of target values $\bar{h}_{i,j}$, defined on a simple background lat-lon grid $\mathcal{G}$. The continuous mesh-spacing function $\bar{h}(\mathbf{x})$ is reconstructed using bilinear interpolation. As will be illustrated in subsequent sections, such an approach provides support for a wide range of mesh-spacing definitions, including those derived from high-resolution topographic data \citep{amante2009etopo1} or solution-adaptive metrics. 

In order to generate high-quality grids, it is necessary to ensure that the imposed mesh-spacing function is \textit{sufficiently-smooth}. Rather than requiring the user to accommodate such constraints, a Lipschitz smoothing process is adopted here. Following the work of \citet{Persson06SizeFunc}, a \textit{gradient-limited} mesh-spacing function $\bar{h}^{'}(\mathbf{x})$ is constructed by constraining the allowable spatial fluctuation over each element in the background grid $\mathcal{G}$. In this study, a scalar smoothing parameter $g\in\mathbb{R}^{+}$ is used to limit variation, such that
\begin{equation}
\qquad \bar{h}^{'}(\mathbf{x}_{i}) \leq \bar{h}^{'}(\mathbf{x}_{j}) + g \cdot \operatorname{dist}\left(\mathbf{x}_{i},\,\mathbf{x}_{j}\right) \,,
\end{equation}
for all adjacent vertex pairs $\mathbf{x}_{i},\,\mathbf{x}_{j}$ in the underlying grid $\mathcal{G}$. The gradient-limited mesh-spacing function $\bar{h}^{'}(\mathbf{x})$ becomes more uniform as $g \rightarrow 0$. In this work, gradient-limiting is implemented following a \textit{fast-marching} method, as described in \citet{Persson06SizeFunc}.

\begin{figure*}
  \centering
  \includegraphics[width=1.\textwidth]{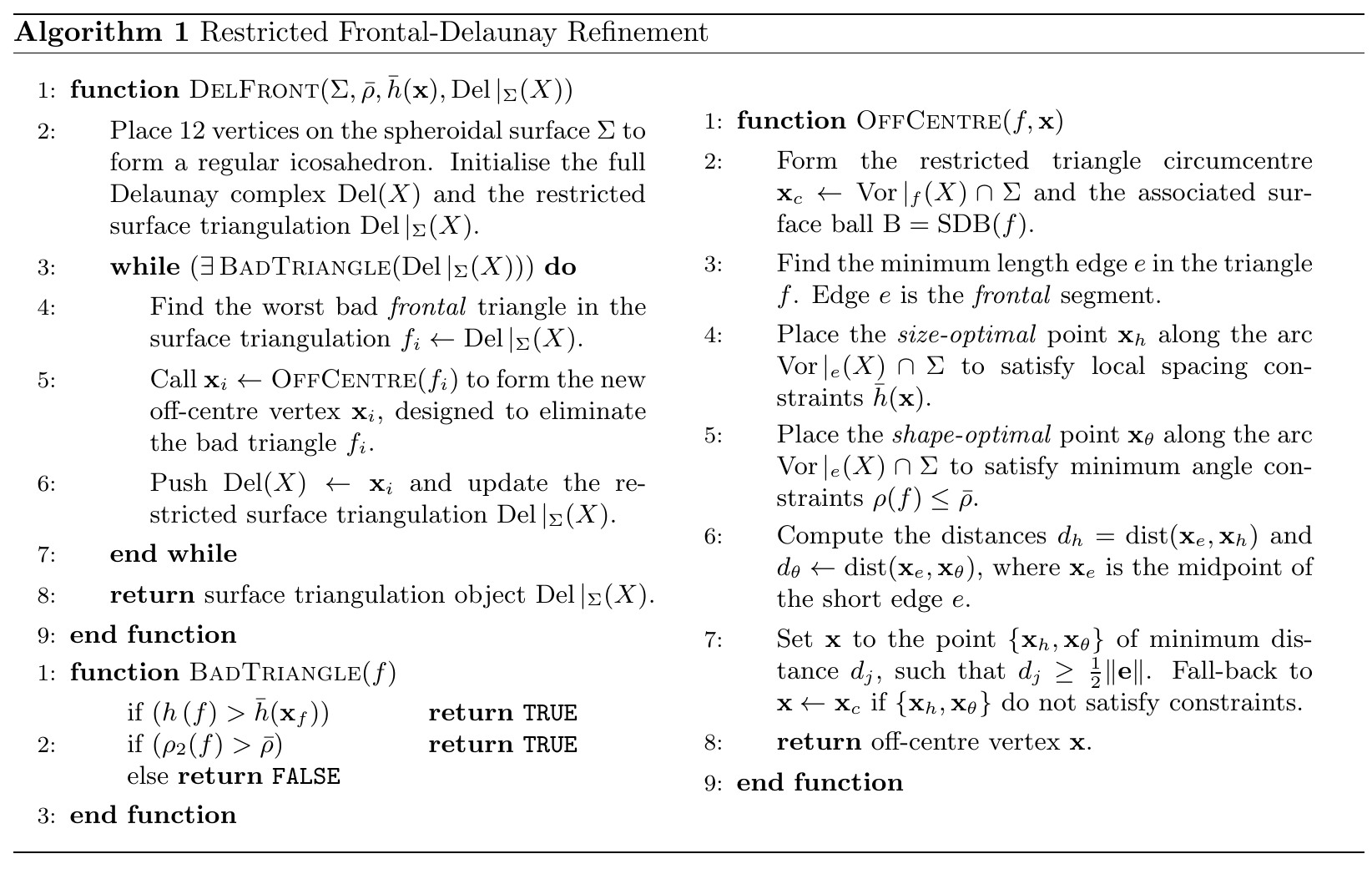}
  
  \label{algorithm_delfront}
\end{figure*}

\subsection{Grid-quality metrics}
\label{section_grid_quality}

The \textit{quality} of a mesh can be evaluated using various grid-quality metrics. Here, a number of standard measures are introduced.

\begin{definition}[radius-edge ratio]
\label{definition_radius_edge}
Given a surface triangle $f_{i}\in\operatorname{Del}|_{\Sigma}(X)$, its \textit{radius-edge} ratio, $\rho(f_{i})$, is given by 
\begin{equation}
\qquad \rho(f_{i}) = \frac{r_{i}}{\|\mathbf{e}_{\text{min}}\|} \,,
\end{equation}
where $r_{i}$ is the radius of the circumscribing ball associated with $f_{i}$ and $\|\mathbf{e}_{\text{min}}\|$ is the length of its shortest edge.
\end{definition}
The radius-edge ratio is a measure of element shape-quality, achieving a minimum of $\tfrac{1}{\sqrt{3}}$ for equilateral triangles and increasing as elements tend toward degeneracy. It is directly related to the minimum angle $\theta_{\text{min}}$ subtended by triangle edges, such that $\theta_{\text{min}} = \operatorname{arcsin}(\frac{1}{2\bar{\rho}})$. Due to the summation of angles in a triangle, this relationship extends to the maximum angle, with $\theta_{\text{max}} \leq \pi - 2 \theta_{\text{min}}$.

\begin{definition}[area-length ratio]
\label{definition_area_length}
Given a surface triangle $f_{i}\in\operatorname{Del}|_{\Sigma}(X)$, its \textit{area-length} ratio, $a(f_{i})$, is given by
\begin{equation}
\qquad a(f_{i}) = \frac{4\sqrt{3}}{3} \, \frac{A_{f}}{\|\mathbf{e}_{\text{rms}}\|^{2}} \,,
\end{equation}
where $A_{f}$ is the signed-area of $f_{i}$ and $\|\mathbf{e}_{\text{rms}}\|$ is the root-mean-square edge length.
\end{definition}
The area-length ratio is a robust, scalar measure of element shape-quality, and is typically normalised to achieve a score of ${+1}$ for ideal elements. The area-length ratio decreases with increasing distortion, achieving a score of ${+0}$ for degenerate elements and ${-1}$ for tangled elements with reversed orientation.

\begin{definition}[relative edge-length]
\label{definition_relative_edge_length}
Given an edge in the surface tessellation $e_{j}\in\operatorname{Del}|_{\Sigma}(X)$, its \textit{relative edge-length}, $h_r(e_{j})$, is given by
\begin{equation}
\qquad h_r(e_{j}) = \frac{\|\mathbf{e}_{j}\|}{\bar{h}(\mathbf{x}_{m})} \,,
\end{equation}
where $\|\mathbf{e}_{j}\|$ is the length of the $j$-th edge and $\bar{h}(\mathbf{x}_{m})$ is the value of the mesh-spacing function sampled at the edge midpoint.
\end{definition}
The relative-length $h_r(e_{j})$ is a measure of mesh-spacing conformance, expressing the ratio of actual-to-desired edge-length. A value of $h_r(e_{j}) = 1$ indicates perfect mesh-spacing conformance.

\subsection{Restricted Frontal-Delaunay refinement}

A high-quality triangular surface mesh is generated on the spheroidal reference surface $\Sigma$ using a \textit{Frontal-Delaunay} variant of the conventional restricted Delaunay-refinement algorithm \citep{boissonnat03ProvablyGoodSurface,boissonnat05ProvablyGoodMeshing,jamin2013cgalmesh,cheng2007sampling,Cheng10PiecewiseSmoothMeshing}. This technique is described by the author in detail in \citet{engwirda2016157,engwirda2016conforming} and differs from standard Delaunay-refinement approaches in terms of the strategies used for the placement of new vertices. Specifically, the Frontal-Delaunay algorithm employs a generalisation of various \textit{off-centre} point-placement techniques \citep{Rebay93FrontalDelaunay,Ungor09OptSteiner}, designed to position vertices such that element-quality and mesh-size constraints are satisfied in a \textit{locally-optimal} fashion. Previous studies have shown that such an approach typically leads to substantial improvements in mean element-quality and mesh smoothness. 

While a variety of solutions for the surface meshing problem have been described previously \citep[e.g.][]{frey2007mesh,rypl1997triangulation,schreiner2006direct,lohner1996regridding}, the restricted Frontal-Delaunay algorithm presented here has been found to offer a unique combination of characteristics --- combining the smooth, high quality grid-generation capabilities of an advancing-front approach, with the theoretical robustness and provable guarantees associated with conventional Delaunay-refinement techniques. Specifically, the Frontal-Delaunay algorithm described here is known to generate grids with very high mean element quality, bounded minimum and maximum angles, tight conformance to grid-spacing constraints, and provable guarantees on topological consistency and convergence. A full description of the algorithm, including detailed discussions of its theoretical foundations and proofs of worst-case grid-quality bounds, can be found in 
\citet{engwirda2016157,engwirda2016conforming}.

\begin{figure}
  \centering
  
  \includegraphics[width=.490\textwidth]{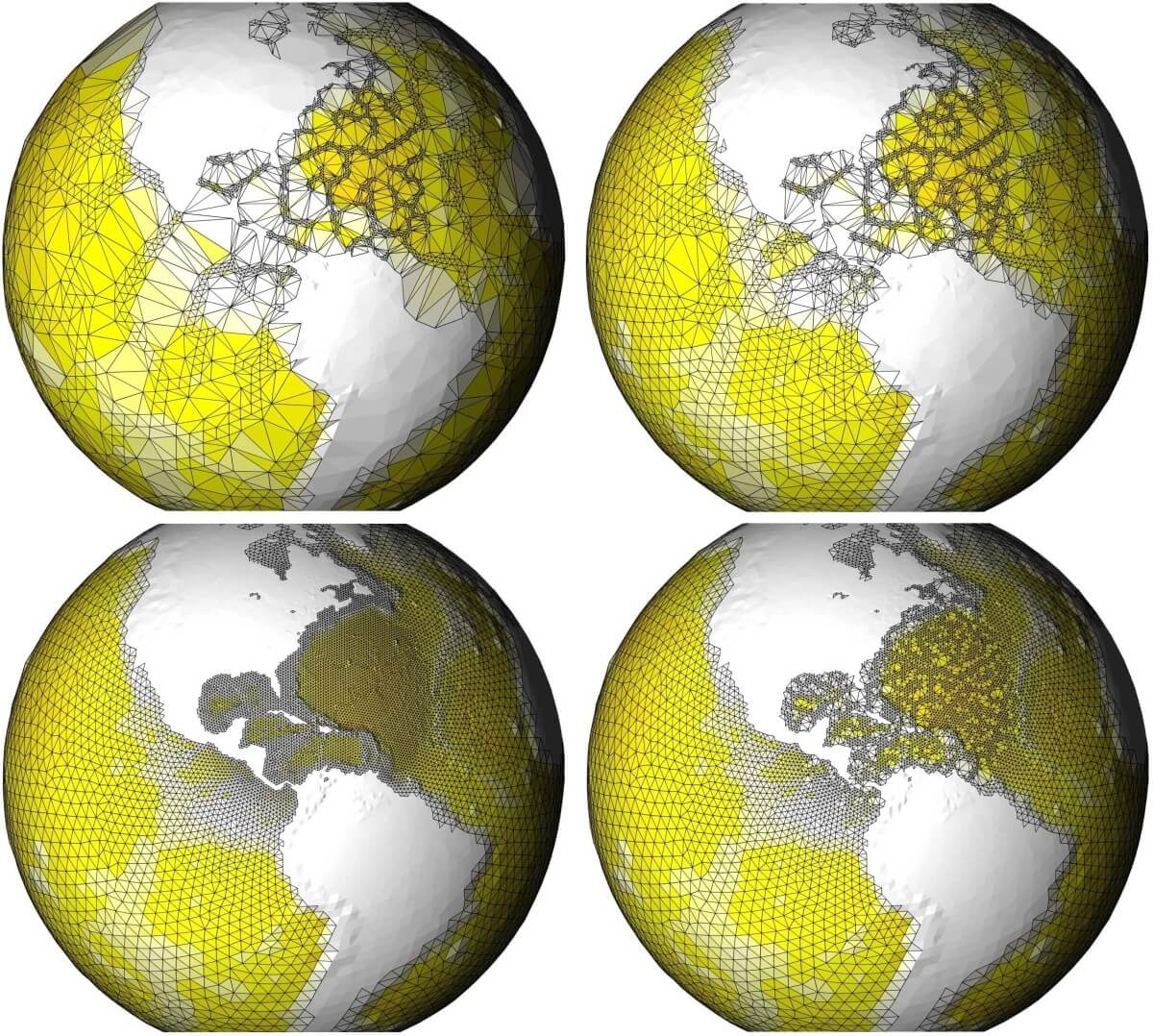}
  
  \caption{Progress of the Frontal-Delaunay refinement algorithm, clockwise from top-left, showing the \textit{space-filling} refinement trajectory.}
  \label{figure_rdel_progress}
\end{figure}

Given a user-defined mesh-spacing function $\bar{h}(\mathbf{x})$ and an upper-bound on the 
element \textit{radius-edge} ratios $\bar{\rho}\geq 1$, the Frontal-Delaunay algorithm
proceeds to sample the spheroidal surface $\Sigma$ by refining any surface triangle that violates either the mesh-spacing or element-quality constraints. The full algorithm is described in Algorithm 1, and detailed snap-shots of the refinement process are shown in Figure~\ref{figure_rdel_progress}. 

Initially, the surface is seeded with a set of twelve vertices to form a standard icosahedron. Refinement then proceeds to insert a new vertex at the \textit{off-centre} refinement point associated with each element to be \textit{eliminated}. Refinement continues until all constraints are satisfied. The refinement process is priority scheduled, with triangles $f_{i}\in\operatorname{Del}|_{\Sigma}(X)$ ordered according to their radius-edge ratios $\rho\left(f_{i}\right)$. This ordering ensures that the element with the \textit{worst} ratio is refined at each iteration. Additionally, triangles are subject to a \textit{frontal} filtering --- requiring that a low-quality element be adjacent to a \textit{converged} triangle before being considered for refinement. This logic helps the algorithm mimic the behaviour of \textit{advancing-front} type techniques, with new vertices and elements expanding from initial seeds. Upon termination, the resulting surface triangulation is guaranteed to contain nicely shaped elements, satisfying both the radius-edge constraints $\rho(f_{i}) \leq \bar{\rho}$ and mesh-spacing bounds $h(f_{i}) \leq \bar{h}(\mathbf{x}_{f})$. Setting $\bar{\rho} = 1$ guarantees that element angles are bounded above $30^\circ$, ensuring that the grid does not contain any highly distorted triangles.

\begin{figure*}
  \centering
 
  \includegraphics[width=.750\textwidth]{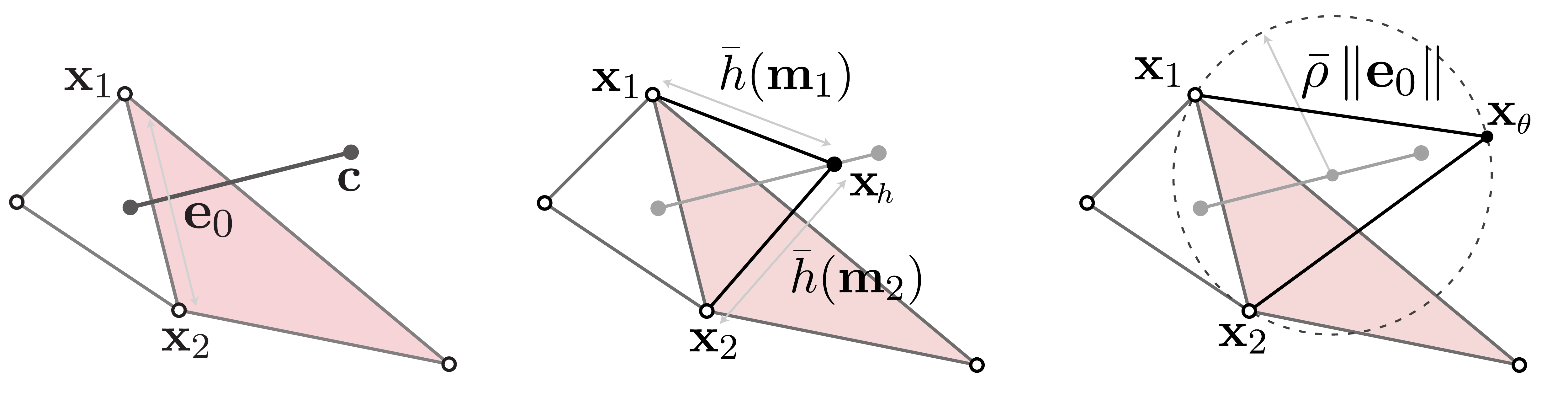}
    
  \caption{Two-dimensional representation of the off-centre point-placement rules, illustrating: (a) a low-quality triangle $f_{i}$ (shaded) to be eliminated, highlighting the \textit{frontal} segment of the Voronoi diagram associated with the short edge $e_{0}\in f_{i}$, (b) placement of the size-optimal point $\mathbf{x}_{h}$ to satisfy local size constraints $\bar{h}(\mathbf{x}_{f})$, and (c) placement of a shape-optimal point $\mathbf{x}_{\theta}$ to satisfy local radius-edge ratios $\bar{\rho}$.}
  \label{figure_offcentres}
\end{figure*}

\subsection{Off-centre point-placement}

The performance of the Frontal-Delaunay algorithm hinges on the use of \textit{off-centre} refinement rules --- locally-optimal point-placement strategies designed to create very high-quality vertex distributions. A set of candidate off-centre points are considered at each new vertex insertion. Type~I vertices, $\mathbf{x}_{c}$, are equivalent to conventional element circumcentres (positioned at the centre of the associated element circumballs), and are used to preserve global convergence. Type~II vertices, $\mathbf{x}_{h}$, are so-called \textit{size-optimal} points, and are designed to satisfy grid-spacing constraints in a locally optimal fashion. Type~III vertices, $\mathbf{x}_{\theta}$, are so-called \textit{shape-optimal} points, and are designed to ensure minimum-angle bounds are satisfied in a worst-first manner. The Type~II and Type~III strategies employed here can be seen as a generalisation of the two-dimensional off-centre techniques presented by \citet{Rebay93FrontalDelaunay} and \citet{Ungor09OptSteiner}, respectively. See Figure~\ref{figure_offcentres} for additional detail.

Given a low-quality triangle $f_{i}\in\operatorname{Del}|_{\Sigma}(X)$ to be \textit{eliminated}, the Type~II and Type~III vertices $\mathbf{x}_{h}$ and $\mathbf{x}_{\theta}$ are positioned along the intersection of an adjacent segment of the Voronoi complex and the spheroidal surface. This intersection $\operatorname{Vor}|_{e}(X)\,\cap\,\Sigma$, defines a \textit{frontal-curve} inscribed on $\Sigma$ --- a \textit{locally-optimal} geodesic segment on which to insert new vertices. Here, $\operatorname{Vor}|_{e}(X)$ is the polygonal face of the Voronoi complex associated with the short edge $e_{0}\in f_{i}$. In the context of conventional advancing-front type methods, the edge $e_{0}$ would denote the current \textit{frontal-segment} about which vertex insertion occurs.

The points $\mathbf{x}_{h}$ and $\mathbf{x}_{\theta}$ are positioned to form new candidate triangles about the frontal edge $e_{0}$, such that local constraints are satisfied optimally. The size-optimal point $\mathbf{x}_{h}$ is positioned to adhere to local grid-spacing constraints, ensuring that new edges satisfy $\|\mathbf{e}_{i}\| \leq \bar{h}(\mathbf{m}_{i})$, where the $\mathbf{m}_{i}$'s are the edge midpoints. These constraints can be solved for an associated altitude 
\begin{gather}
\qquad a_{h} = \min \bigg(\tilde{a}_{h},\, \tfrac{1}{2}\sqrt{3}\,\tilde{h}\bigg)\,, 
\quad \text{where} 
\\[1ex]
\qquad \tilde{a}_{h} = \sqrt{\tilde{h}^{2} - \tfrac{1}{2}\|\mathbf{e}_{0}\|^{2}}\,,
\quad \tilde{h} = \frac{\bar{h}(\mathbf{m}_{1})+\bar{h}(\mathbf{m}_{2})}{2} \,.
\end{gather}

Similarly, the shape-optimal point is positioned to adhere to minimum angle constraints, sliding $\mathbf{x}_{\theta}$ along the inscribed curve $\operatorname{Vor}|_{e}(X)\,\cap\,\Sigma$ so that the new candidate triangle satisfies $\rho \leq \bar{\rho}$. These constraints lead to an associated altitude
\begin{gather}
\qquad a_{\theta} = \frac{\|\mathbf{e}_{0}\|}{\tilde{\beta}_{\theta}} \,, 
\quad \text{where}
\\[1ex]
\qquad \tilde{\beta}_{\theta} = 2\, \tan\bigg(\frac{\tilde{\theta}}{2}\bigg) \,,
\quad \tilde{\theta} = \operatorname{arcsin}\bigg(\frac{1}{2\, \bar{\rho}}\bigg) \,.
\end{gather}

Given $a_{h}$ and $a_{\theta}$, the position of the points $\mathbf{x}_{h}$ and $\mathbf{x}_{\theta}$ is calculated by computing the intersection of balls of radius $a_{h}$ and $a_{\theta}$, centred at the midpoint of the frontal edge $e_{0}$ and the \textit{frontal} curve $\operatorname{Vor}|_{e}(X)\,\cap\,\Sigma$. This approach ensures that new vertices are positioned by advancing a specified distance along the surface $\Sigma$ in the \textit{frontal} direction. For non-uniform $\bar{h}(\mathbf{x})$, expressions for the position of the point $\mathbf{x}_{h}$ are non-linear, with the altitude $a_{h}$ depending on an evaluation of the mesh-size function at the edge midpoints $\bar{h}(\mathbf{m}_{i})$ and visa-versa. In practice, since $\bar{h}(\mathbf{x})$ is guaranteed to be smooth, a simple iterative predictor-corrector procedure is sufficient to solve these expressions approximately.

Finally, given the set of candidate off-centre vertices $\left\{\mathbf{x}_{c},\,\mathbf{x}_{h},\,\mathbf{x}_{\theta}\right\}$, the position of the refinement point $\mathbf{x}$ for the triangle $f_{i}$ is selected. A \textit{worst-first} strategy is adopted, choosing the point that satisfies local constraints in a greedy fashion. Specifically, the closest point lying on the adjacent Voronoi segment $\operatorname{Vor}|_{e}(X)$ and outside the neighbourhood of the frontal edge $e_{0}$ is selected, with
\begin{gather}
\label{eqn_point_selection}
\qquad \mathbf{x} = \left\{
\begin{array}{ll}
\mathbf{x}_{j}, & 
\text{ if }\bigg(d_{j} \leq d_{c} \text{ and } d_{j} \geq \tfrac{1}{2}\|\mathbf{e}_{0}\| \bigg)
\\[2ex]
\mathbf{x}_{c}, &
\text{ otherwise}
\end{array}
\right. ,
\end{gather}

where $j = \operatorname{argmin}\left(d_{h},\, d_{\theta}\right)$ and $d_{j} = \operatorname{dist}\left(\mathbf{x}_{e},\mathbf{x}_{j}\right)$ are distances from the midpoint of the frontal edge $e_{0}$ to the size- and shape-optimal points $\mathbf{x}_{h}$ and $\mathbf{x}_{\theta}$, respectively. This cascading selection criteria seeks a balance between local optimality and global convergence --- smoothly degenerating to a conventional circumcentre-based Delaunay-refinement strategy in limiting cases, while using locally optimal points where possible. Specifically, these constraints guarantee that refinement points always lie within a local \textit{safe} region on the Voronoi complex --- being positioned on an adjacent Voronoi segment and bound between the circumcentre of the element itself and the diametric ball of the associated frontal edge. This condition ensures that new points are never positioned too close to an existing vertex, leading to provable guarantees on the performance of the algorithm. See previous work by the author \citep{engwirda2016157,engwirda2016conforming} for additional detail.

\subsection{Additional remarks}

As a \textit{restricted} Delaunay-refinement approach, a full three-dimensional Delaunay tetrahedralisation $\operatorname{Del}(X)$ is incrementally maintained throughout the surface meshing phase. As per Definition~\ref{definition_restricted_tessellation}, the set of restricted surface triangles $\operatorname{Del}|_{\Sigma}(X)$ that lie on the spheroid $\Sigma$, are expressed as a subset of the faces of this tetrahedral complex, such that $\operatorname{Del}|_{\Sigma}(X)\subseteq\operatorname{Del}(X)$. In an effort to minimise the expense of maintaining this full-dimensional tessellation, an additional \textit{scaffolding} vertex $\mathbf{x}_{s}$ is initially inserted at the centre of the spheroid. This has the effect of simplifying the topological structure of $\operatorname{Del}(X)$, with the resulting tetrahedral elements forming a simple \textit{wheel-like} configuration --- emanating radially outward from the scaffolding vertex $\mathbf{x}_{s}$. See Figure~\ref{figure_restricted_faces} for additional detail.

As shown in Figure~\ref{figure_rdel_progress}, one interesting characteristic of the Frontal-Delaunay algorithm described here relates to the \textit{refinement-trajectory} that is followed when inserting new vertices and triangles. Unlike standard Delaunay-refinement or advancing-front type methods, it can be seen that the algorithm adopts a \textit{space-filling curve} type pattern, covering the surface in a fractal-like configuration, before recursively filling in the gaps. Note that no explicit space-filling curve constraint has been implemented here --- this behaviour is simply an emergent property of the algorithm itself, and is due to interactions between the greedy priority schedule, the frontal filtering and the off-centre point-placement strategies. In practice, it has been found that this space-filling type behaviour leads to the construction of very high-quality triangulations, typically exceeding the performance of standard advancing-front or Delaunay-refinement type schemes.

\section{Hill-climbing mesh optimisation}
\label{section_optimisation_algorithm}

\begin{figure*}
  \centering
  \includegraphics[width=1.\textwidth]{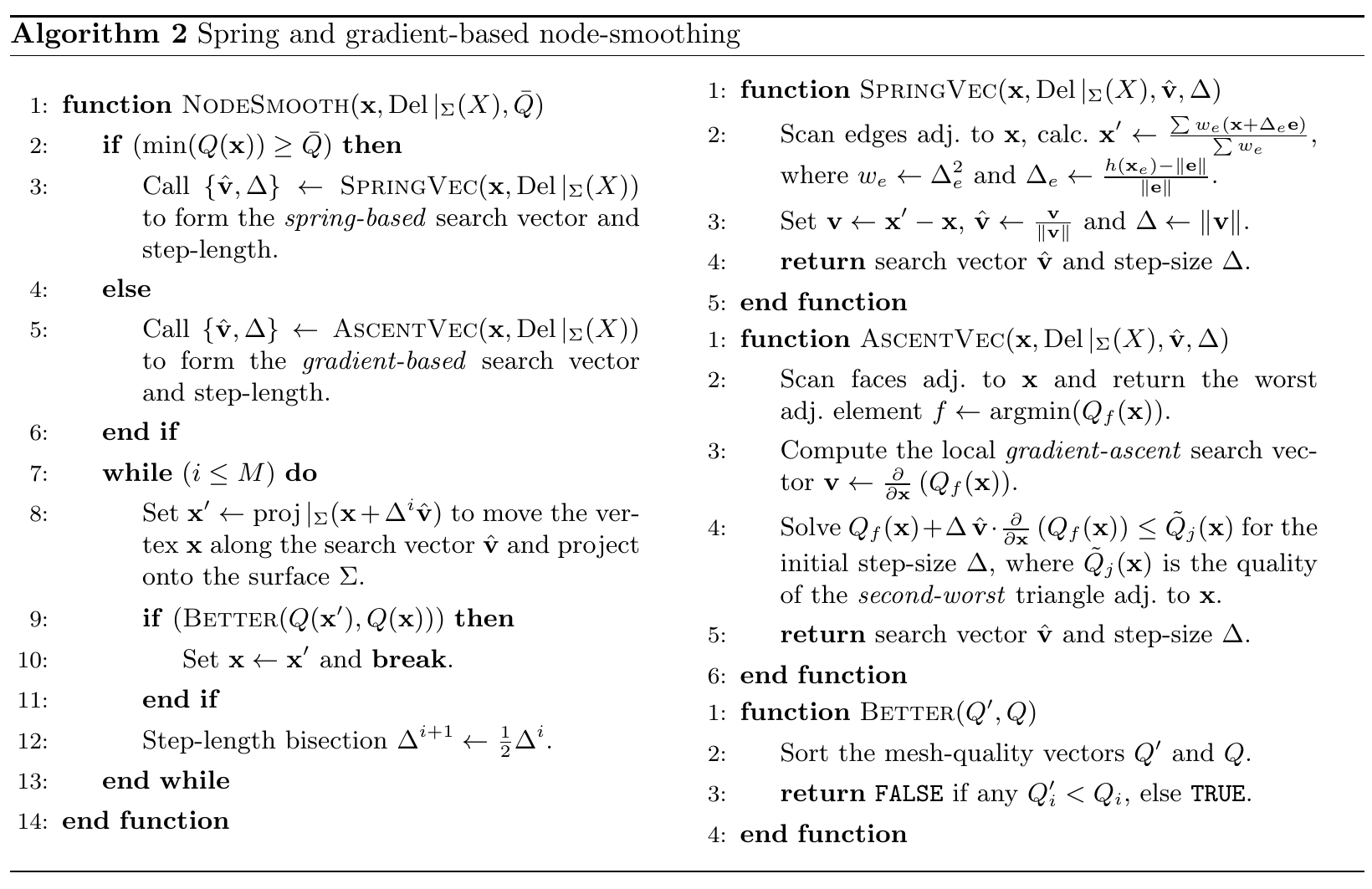}
  
  \label{algorithm_node_smooth}
\end{figure*}

While the Delaunay grids produced by the refinement algorithm described previously are guaranteed to be of very high-quality, producing triangulations with angles bounded between $\theta_{f}=30^\circ$ and $\theta_{f}=120^\circ$, these tessellations can often be further improved through subsequent \textit{mesh-optimisation} operations. Recalling that the construction of \textit{well-centred} grids appropriate for unstructured C-grid schemes \textit{require} that maximum angles be bounded below $\theta_{f}=90^\circ$, the application of such optimisation procedures can in fact be seen as a \textit{necessary} component of the grid-generation work-flow explored here.

In the present work, grid-optimisation is realised as a coupled geometrical and topological optimisation task --- seeking to reposition vertices and update grid topology to maximise a given element-wise mesh-quality metric $\mathcal{Q}_{f}(\mathbf{x})$. A \textit{hill-climbing} type optimisation strategy is pursued, in which a locally-optimal solution is sought based on an initial grid configuration. In this study, the grid is optimised according to the \textit{area-length} quality metric (see Definition~\ref{definition_area_length}), a robust scalar measure of grid-quality that achieves a score of ${+1}$ for \textit{perfect} elements --- decreasing toward zero with increasing levels of element distortion. Optimisation predicates are implemented following a \textit{hill-climbing} strategy \citep{freitag1997tetrahedral,klingner2008aggressive}, with modifications to the grid accepted only if the local mesh-quality metrics are sufficiently improved. Specifically, a \textit{worst-first} strategy is adopted here, in which each given operation is required to improve the \textit{worst-case} quality associated with elements in the local subset acted upon. Such a philosophy ensures that global mesh-quality is increased monotonically as optimisation proceeds. Note that such behaviour is designed to maximise the quality of the \textit{worst} elements in the grid, rather than improving a mean measure. This represents an important distinction when compared to other iterative mesh-optimisation algorithms, such as Centroidal Voronoi Tessellation (CVT) type schemes \citep{jacobsen2013parallel}, in which all entities in the grid are adjusted simultaneously until a global convergence criterion is satisfied.

\subsection{`Spring'-based mesh smoothing}

Considering firstly the geometric optimality of the grid, a \textit{mesh-smoothing} procedure is undertaken --- seeking to reposition the nodes of the grid to improve element quality and mesh-spacing conformance. Following the work of \citet{persson2004simple}, a \textit{spring-based} approach is pursued, in which edges in the Delaunay triangulation are treated as \textit{elastic-rods} with a prescribed natural length. Nodes are iteratively repositioned until a local equilibrium configuration is reached. See Algorithm 2 for full details. In the original work of Persson, nodal positions are adjusted via a local time-stepping loop, with all nodes updated concurrently under the action of explicit spring forces. In the current study, a non-iterative variant is employed, in which each node is repositioned one-by-one, such that constraints in each local neighbourhood are satisfied directly. Specifically, a given node $\mathbf{x}_{i}$ is repositioned as a weighted sum of contributions from incident edges
\begin{gather}
\label{equation_springs}
\qquad \mathbf{x}_{i}^{n+1} = \frac{\sum w_{k} \left(\mathbf{x}_{i}^{n} + \Delta_{k}\mathbf{v}_{k}\right) }{\sum w_{k}} \,,
\quad \text{where}
\\[1ex]
\qquad \mathbf{v}_{k} = \mathbf{x}_{i}^{n}-\mathbf{x}_{j}^{n} 
\quad\text{and}
\quad \Delta_{k} = \frac{\bar{h}(\mathbf{x}_{k}^{n}) - l_{k}}{l_{k}} \,.
\end{gather}

Here, $\mathbf{x}_{i}^{n},\,\mathbf{x}_{j}^{n}$ are the current positions of the two nodes associated with the $k$-th edge, $l_{k}$ is the edge length and $\bar{h}(\mathbf{x}_{k})$ is the value of the mesh-spacing function evaluated, at the edge midpoint. $\Delta_{k}$ is the relative spring \textit{extension} required to achieve equilibrium in the $k$-th edge. The scalars $w_{k}\in\mathbb{R}^{+}$ are edge weights. Setting $w_{k}=1$ results in an unweighted scheme, consisting of simple \textit{linear} springs. In this study, the use of nonlinear weights, defined by setting $w_{k} = \Delta_{k}^{2}$, was found to offer superior performance.

Noting that application of the spring-based operator (\ref{equation_springs}) may move nodes away from the underlying spheroidal surface $\Sigma$, an additional \textit{projection} operator is introduced to ensure that the grid conforms to the surface geometry exactly. Following the application of each spring-based adjustment, nodes are moved back onto the geometry via a closest-point projection.  

Consistent with the hill-climbing paradigm described previously, each nodal adjustment is required to be \textit{validated} before being committed to the updated grid configuration. Specifically, nodal adjustments are accepted only if there is sufficient improvement in the mesh-quality metrics associated with the set of adjacent elements. A sorted comparison of quality metrics before and after nodal repositioning is performed, with nodal adjustments successful if grid-quality is improved in a \textit{worst-first} manner. This \textit{lexicographical} quality comparison is consistent with the methodology employed in \citet{klingner2008aggressive}.

\subsection{Gradient-based mesh smoothing}

While the spring-based smoothing operator is effective in adjusting a grid to satisfy mesh-spacing constraints, and tends to improve the quality of the grid on average, it is not guaranteed to improve element quality in all cases. As such, an additional \textit{steepest-ascent} strategy is pursued \citep{freitag1997tetrahedral}, in which nodal positions are adjusted based on the local gradients of incident element quality functions. See Algorithm 2 for full details. Specifically, a given node $\mathbf{x}_{i}$ is repositioned along a local \textit{search-vector} chosen to improve the quality of the worst incident element 
\begin{gather}
\label{equation_grad_ascent}
\qquad \mathbf{x}_{i}^{n+1} = \mathbf{x}_{i}^{n} + \Delta_{f} \hat{\mathbf{v}}_{f} \,,
\quad \text{where}
\\[1ex]
\qquad \mathbf{v}_{f} = \frac{\partial}{\partial\mathbf{x}} \left( \mathcal{Q}_{f}(\mathbf{x}_{i}) \right) \,,
\quad f = \operatorname{arg\,min}_{j}\left(\mathcal{Q}_{j}(\mathbf{x}_{i})\right) \,.
\end{gather}

Here, the index $j$ is taken as a loop over the Delaunay triangles $f_{j}\in\operatorname{Del}|_{\Sigma}(X)$ incident to the node $\mathbf{x}_{i}$. The scalar step-length $\Delta_{f} \in \mathbb{R}^{+}$ is computed via a line-search along the gradient ascent vector $\hat{\mathbf{v}}_{f}$, and in this study, is taken as the first value that leads to an improvement in the worst-case incident quality metric $\mathcal{Q}_{f}(\mathbf{x}_{i})$. This length is computed by a simple bisection strategy, iteratively testing $\Delta_{f}^{p} = \left(\frac{1}{2}\right)^{p} \alpha$ until a successful nodal adjustment is found. Here $p$ denotes the local line-search iteration, while the scalar length $\alpha \in \mathbb{R}^{+}$ is determined as a solution to the first-order Taylor expansion 
\begin{equation}
\qquad \mathcal{Q}_{f}(\mathbf{x}_{i}) + \alpha\, \hat{\mathbf{v}}_{f} \cdot \frac{\partial}{\partial\mathbf{x}} \left( \mathcal{Q}_{f}(\mathbf{x}_{i}) \right) \leq \tilde{\mathcal{Q}}_{j}(\mathbf{x}_{i}) \,.
\end{equation}

The quantity $\tilde{\mathcal{Q}}_{j}(\mathbf{x}_{i})$ is the \textit{second-lowest} grid-quality score associated with the set of triangles $f_{j}\in\operatorname{Del}|_{\Sigma}(X)$ incident to the central node $\mathbf{x}_{i}$. This selection strategy \citep{freitag1997tetrahedral} is designed to compute an initial displacement $\alpha$ that will improve the worst element in the set until its quality is equal that of its next best neighbour. Noting that such an expansion is only first-order accurate, the step-length is iteratively decreased using the bisection strategy described previously. A limited line-search is employed here, testing iterations $p \leq 5$ until a successful step is found. Consistent with the spring-based procedure described previously, a geometry-projection operator is implicitly incorporated within each update (\ref{equation_grad_ascent}), ensuring that nodes remain constrained to the spheroidal surface.

\begin{figure}
  \centering
  
  \includegraphics[width=.480\textwidth]{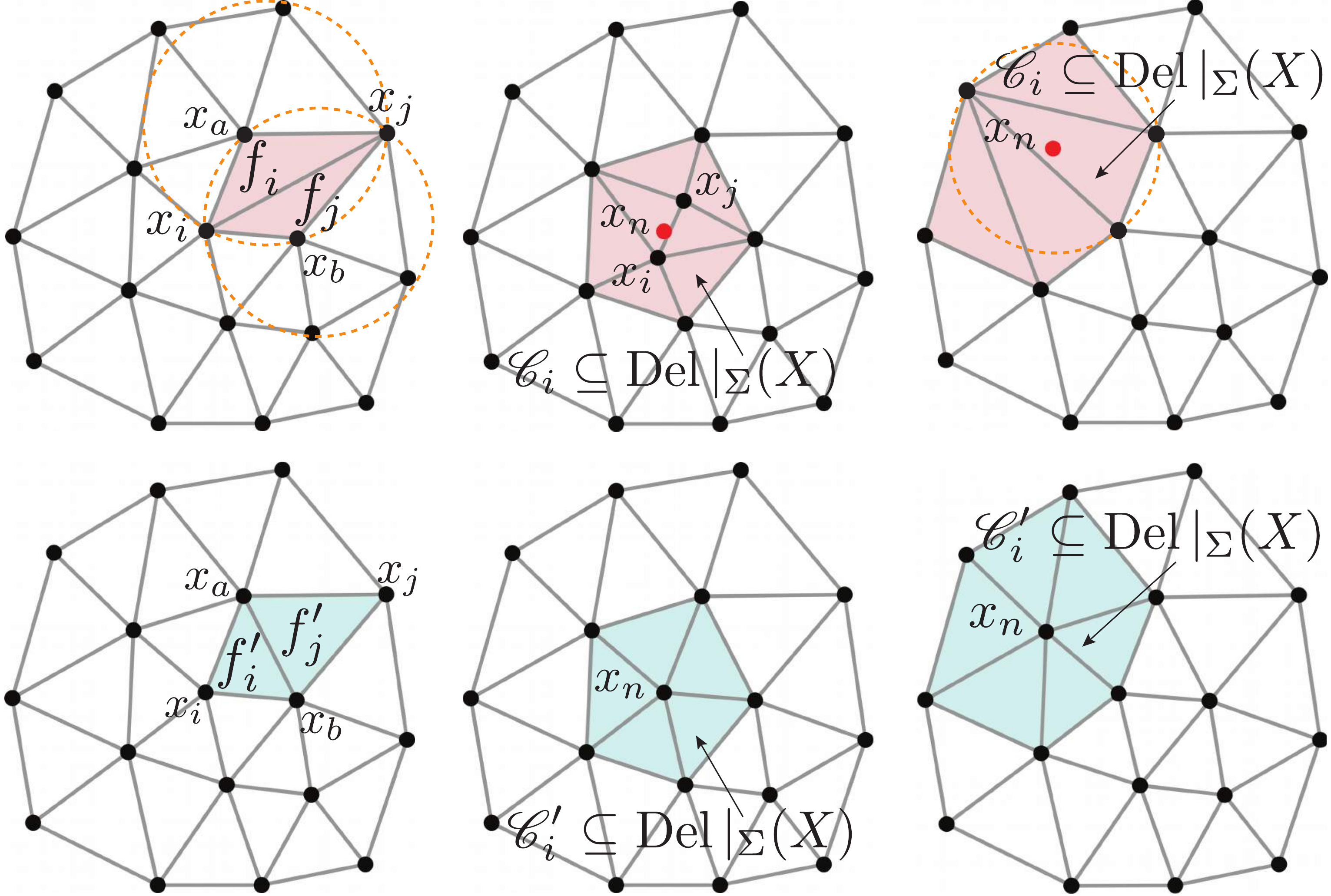}
  
  \caption{Topological operations for grid optimsiation, showing (left) an edge-flip, (middle) an edge-contraction, and (c) an edge-refinement operation. Grid configurations before and after each flip are shown in the upper and lower panels, respectively.}
  \label{figure_topo_flip}
\end{figure}

\subsection{Topological `flips'}

In addition to purely geometrical operations, grid optimisation also requires that adjustments be made to the underlying mesh topology, such that the surface triangulation remains a valid Delaunay structure. While it is possible to simply re-compute the full Delaunay tessellation after each adjustment, such an approach would impose significant computational costs, especially when considering that a majority of updates involve small perturbations. In this work, an alternative strategy is pursued, based on local element-wise transformations, known as topological \textit{flips}. 

For any given pair of adjacent surface triangles $\{f_{i},\,f_{j}\}\in\operatorname{Del}|_{\Sigma}(X)$, a local re-triangulation can be achieved by \textit{flipping} the local connectivity about the shared edge $\{x_i,\,x_j\}$, forming a new edge between the opposing vertices $\{x_a,\,x_b\}$. Such an operation results in the deletion of the existing triangles $\{f_{i},\,f_{j}\}$ and the creation of a new pair $\{f_{i}',\,f_{j}'\}$. This operation is illustrated in Figure~\ref{figure_topo_flip}a. In the present study, the iterative application of such \textit{edge-flipping} operations is used to adjust the topology of the surface triangulation, ensuring that it remains Delaunay throughout the optimisation phase. Specifically, given a general, possibly non-Delaunay, surface triangulation $\operatorname{Tri}|_{\Sigma}(X)$, a cascade of edge-flips are used to reach a valid restricted Delaunay surface tessellation $\operatorname{Del}|_{\Sigma}(X)$. For each adjacent triangle pair $\{f_{i},\,f_{j}\}\in\operatorname{Tri}|_{\Sigma}(X)$ an edge-flip is undertaken if a local violation of the Delaunay criterion is detected. New elements created by successful edge-flips are iteratively re-examined until no further modifications are necessary. This approach follows the standard flip-based algorithms described in, for instance \citet{lawson77flips,ChengDeyShewchuk}. 

Given a triangle $f_{i}\in\operatorname{Tri}|_{\Sigma}(X)$, the local Delaunay criterion is violated if there exists a node $x_{q} \notin f_{i}$ \textit{interior} to the associated circumscribing ball. In this work, such violations are detected by considering the \textit{restricted} circumballs $\operatorname{B}(\mathbf{c}_{i},r_{i})$ associated with each triangle $f_{i}\in\operatorname{Tri}|_{\Sigma}(X)$, where the centre of the balls $\mathbf{c}_{i}$ are a projection of the \textit{planar} element circumcentres onto the spheroidal surface $\Sigma$. Such constructions are designed to account for the curvature of the underlying surface. Given an adjacent triangle pair $\{f_{i},\,f_{j}\}\in\operatorname{Tri}|_{\Sigma}(X)$ an edge-flip is undertaken if either opposing vertex $x_a,\,x_b$ lies within the circumball associated with the adjacent triangle. To prevent issues associated with exact floating-point comparisons, a small relative tolerance is incorporated. Specifically, nodes are required to penetrate the opposing circumball by a distance $\epsilon$ before an edge-flip is undertaken, with $\epsilon = \tfrac{1}{2}\left(r_{i}+r_{j}\right)\bar{\epsilon}$ and $\bar{\epsilon} = 1 \times 10^{-10}$ in the current double-precision implementation.

\begin{figure*}
  \centering
  \includegraphics[width=1.\textwidth]{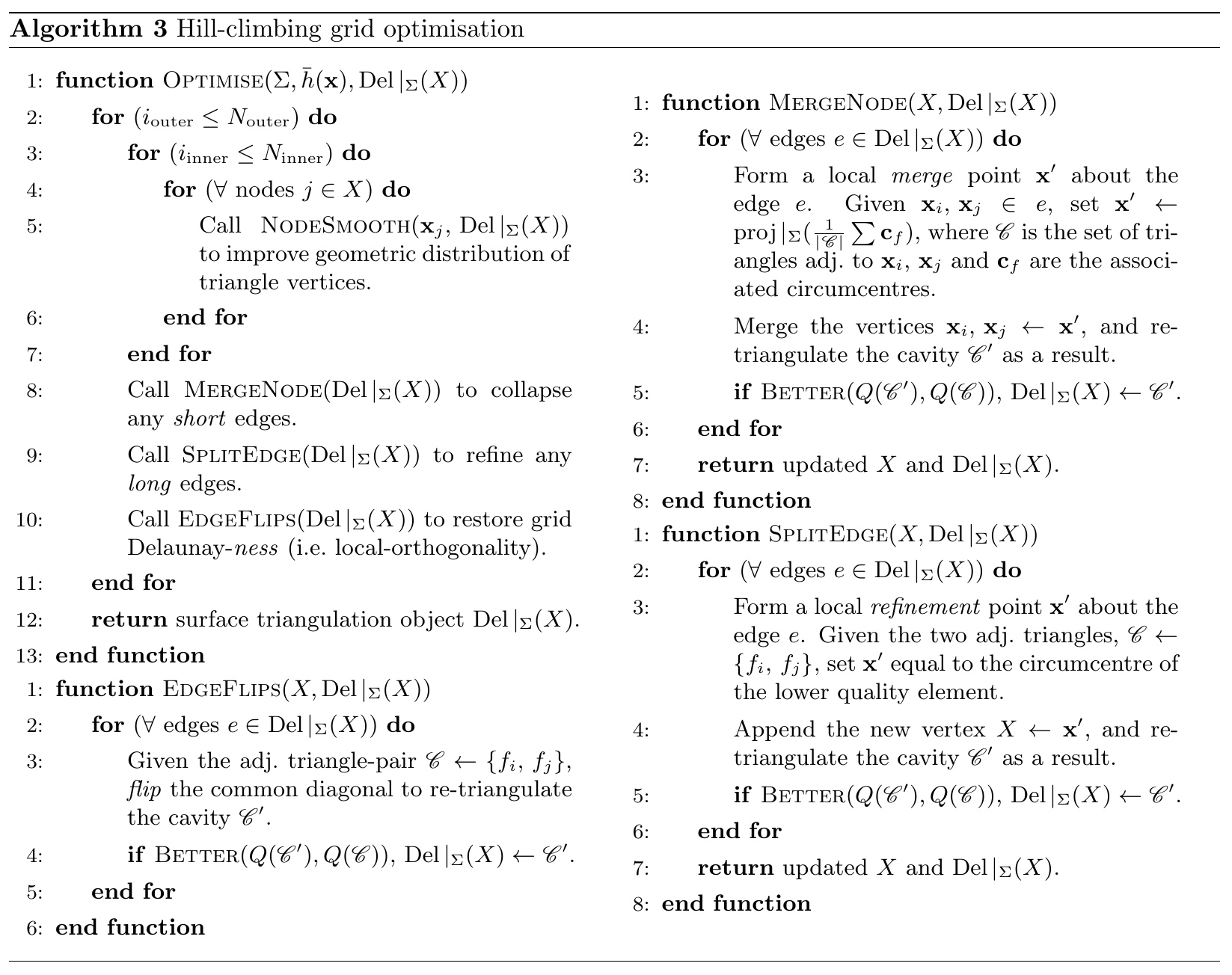}
  
  \label{algorithm_hill_climb}
\end{figure*}

\subsection{Edge contraction}

In some cases, grid-quality and mesh-spacing conformance can be improved through the application of so-called \textit{edge-contraction} operations, whereby nodes are removed from the grid by collapsing certain edges. Given an edge $e_{k}\in\operatorname{Tri}|_{\Sigma}(X)$, a re-triangulation of the local \textit{cavity} $\mathcal{C}_{i}\subseteq\operatorname{Tri}|_{\Sigma}(X)$, formed by the set of triangles incident to the nodes $\{x_i,\,x_j\}\in e_{k}$, can be achieved by \textit{merging} the nodes $x_i$ and $x_j$ at some local mean position. In addition to collapsing the edge $e_{k}$, edge-contraction also removes the two surface triangles $\{f_{i},\,f_{j}\}\in\operatorname{Tri}|_{\Sigma}(X)$ adjacent to $e_{k}$, resulting in a re-triangulation of the local cavity $\mathcal{C}_{i}'\subseteq\operatorname{Tri}|_{\Sigma}(X)$. See Figure~\ref{figure_topo_flip}b for illustration. In the present work, nodes are merged to a mean position $\mathbf{x}_{n}$ --- taken as the average of adjacent element circumcentres, such that $\mathbf{x}_{n} = \tfrac{1}{|\mathcal{C}_{i}|}\sum \mathbf{c}_{j}$. Here, the $\mathbf{c}_{j}$'s are the centres of the circumballs associated with the triangles $f_{j}\in\mathcal{C}_{i}$. As per previous discussions, the mean position $\mathbf{x}_{n}$ is projected onto the spheroidal surface $\Sigma$. While such an approach is slightly more computationally intensive than strategies based on collapsing to edge-midpoints, this weighted circumcentre-based scheme was found to be substantially more effective in practice. Consistent with the hill-climbing philosophy pursued throughout this study, edge-contraction operations are only utilised if there is sufficient improvement in local mesh-quality metrics. Specifically, edge-contraction is undertaken if the quality of the new cavity $\mathcal{C}_{i}'$ exceeds that of the initial configuration $\mathcal{C}_{i}$ in a worst-first fashion.

\subsection{Edge refinement}

Fulfilling the opposite role to edge-contraction, so-called \textit{edge-refinement} operations seek to improve the grid through the addition of new nodes and elements. In the present study, a simplified refinement operation is utilised, in which a given edge $e_{k}\in\operatorname{Tri}|_{\Sigma}(X)$ is refined by placing a new node $x_{n}$ at the centre of the restricted circumball $\operatorname{B}(\mathbf{c}_{i},r_{i})$ associated with the lower quality adjacent triangle $f_{i}\in\operatorname{Tri}|_{\Sigma}(X)$. Insertion of the new node $x_{n}$ induces a re-triangulation of the local cavity $\mathcal{C}_{i}\in\operatorname{Tri}|_{\Sigma}(X)$ --- constructed by expanding about $x_{n}$ in a local greedy fashion. Starting from the initial cavity $\mathcal{C}_{i} = \{f_{i},\,f_{j}\}$, where $\{f_{i},\,f_{j}\}\in\operatorname{Tri}|_{\Sigma}(X)$ are the triangles adjacent to the edge $e_{k}$, additional elements are added in a \textit{breadth-first} manner, with a new, unvisited neighbour $f_{k}$ added to the cavity $\mathcal{C}_{i}$ if doing so will improve the worst-case element quality metric. The final cavity $\mathcal{C}_{i}$ is therefore a locally-optimal configuration. In practice, this iterative deepening of $\mathcal{C}_{i}$ typically convergences in one or two iterations. See Figure~\ref{figure_topo_flip}c for illustration. As per the edge-contraction and node-smoothing operations described previously, edge-refinement is implemented according to a hill-climbing type philosophy, with operations successful only if there is sufficient improvement in local grid-quality. Consistent with previous discussions, a lexicographical comparison of the grid-quality metrics associated with elements in the initial and final states $\mathcal{C}_{i}$ and $\mathcal{C}_{i}'$ is used to determine success.

\subsection{Optimisation schedule}

The full grid-optimisation procedure is realised as a combination of the various geometrical and topological operations described previously, organised into a particular iterative optimisation \textit{schedule}. See Algorithm 3 for full details. Each outer iteration consists of a fixed set of operations: four node-smoothing passes, a single pass of edge refinement/contraction operations, and, finally, iterative edge-flipping to restore the Delaunay criterion. In this study, sixteen outer iterations are employed. Each node-smoothing pass is a composite operation, with the spring-based technique used to adjust nodes adjacent to high-quality elements, and the gradient-ascent method used otherwise. Specifically, spring-based smoothing is used to adjust nodes adjacent to elements with a minimum quality score of $\bar{\mathcal{Q}}_{f} \geq 0.9375$. Such thresholding ensures that the expensive gradient-ascent type iteration is reserved for the worst elements in the grid. The optimisation schedule employed here is not based on any rigorous theoretical derivation, but is simply a set of heuristic choices that have proven to be effective in practice. The application of multiple node-smoothing passes within an outer iteration containing subsequent topological, contraction and refinement type operations is consistent with the methodologies employed in, for instance \citet{freitag1997tetrahedral,klingner2008aggressive}.

\section{Results \& Discussions}
\label{section_results}

\begin{figure*}
  \centering

  \includegraphics[width=.975\textwidth]{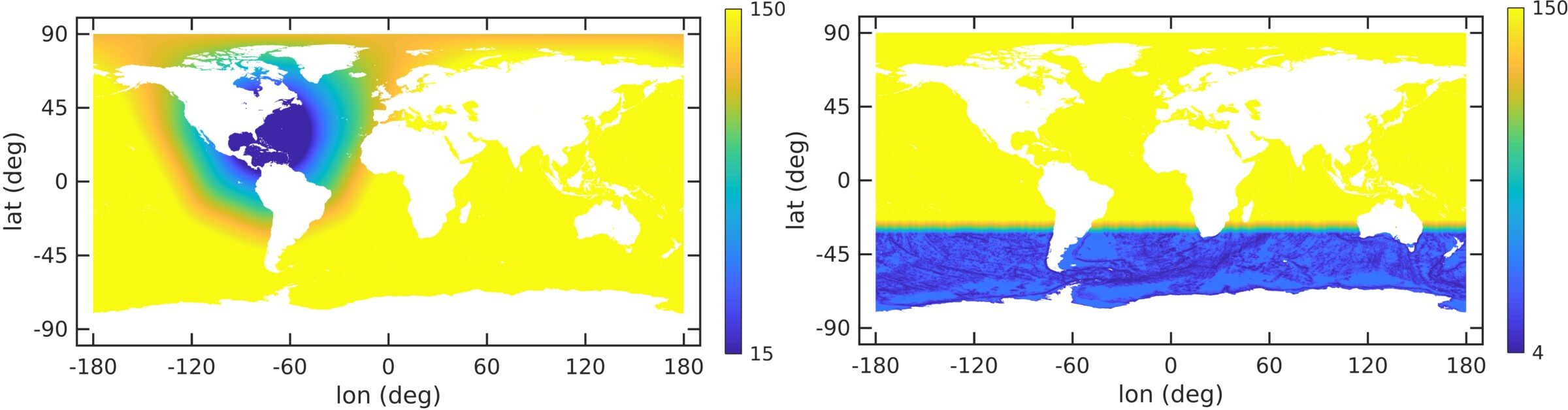}
  
  \caption{Mesh-spacing functions $\bar{h}(\mathbf{x})$ for the regionally-refined North Atlantic and topographically-refined Southern Ocean grids. Mesh-spacing is shown in km. }
  \label{figure_grid_spac}
\end{figure*}

The performance of the Frontal-Delaunay refinement and hill-climbing optimisation algorithms presented in Sections~\ref{section_refinement_algorithm} and \ref{section_optimisation_algorithm} was investigated experimentally, with the methods used to mesh a series of benchmark problems. The algorithm was implemented in {\nolinebreak C\texttt{++}} and compiled as a 64-bit executable. The full algorithm has been implemented as a specialised variant of the general-purpose JIGSAW meshing package, denoted JIGSAW-GEO, and is currently available online \citep{JIGSAWGEOpackage} or by request from the author. All tests were completed on a Linux platform using a single core of an Intel i7 processor. Visualisation and post-processing was completed using MATLAB.

\subsection{Preliminaries}

\begin{figure*}
  \centering
  \includegraphics[width=.9675\textwidth]{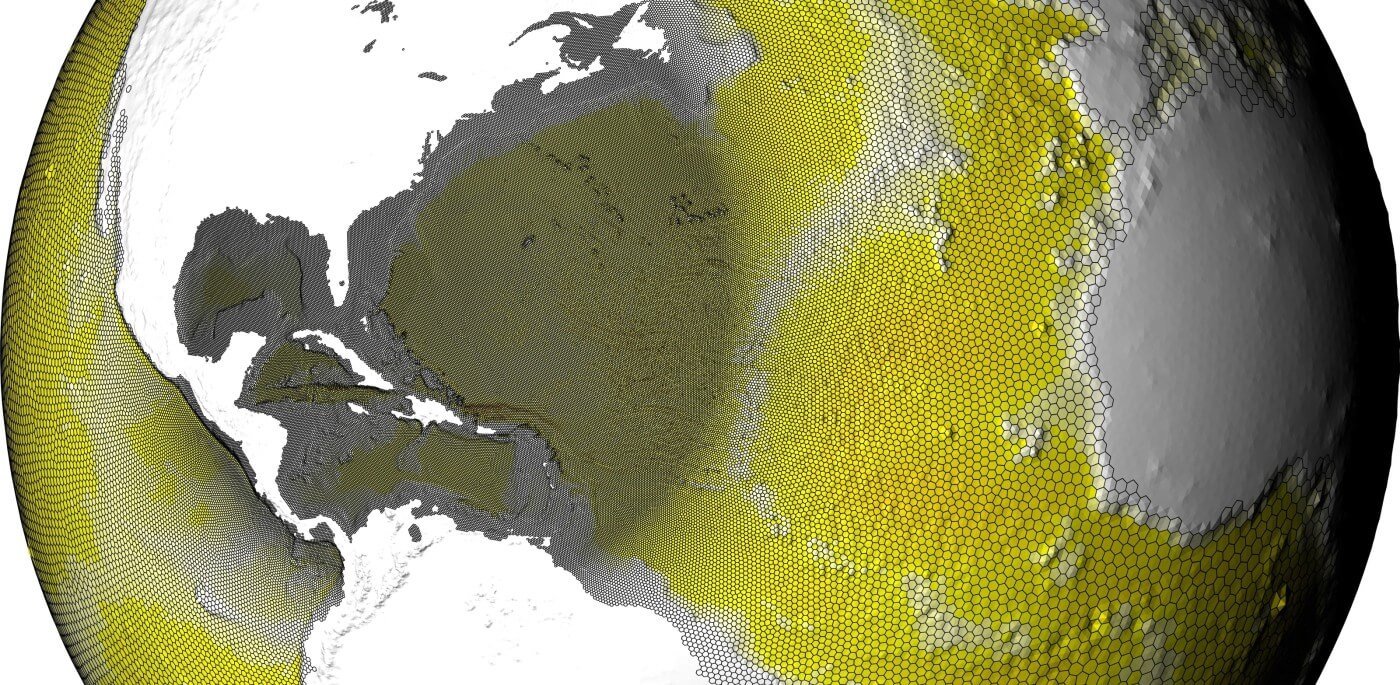}
  
  \caption{A regionally-refined Voronoi-type grid of the North Atlantic region. Global coarse grid resolution is $150\,\mathrm{km}$, with a $15\,\mathrm{km}$ eddy-permitting grid-spacing specified over the Atlantic ocean basin. Topography is drawn using an exaggerated scale, with elevation from the reference geoid amplified by a factor of 10.}
  \label{figure_regional_grid}
\end{figure*}

\begin{figure*}
  \centering
  
  \includegraphics[width=.865\textwidth]{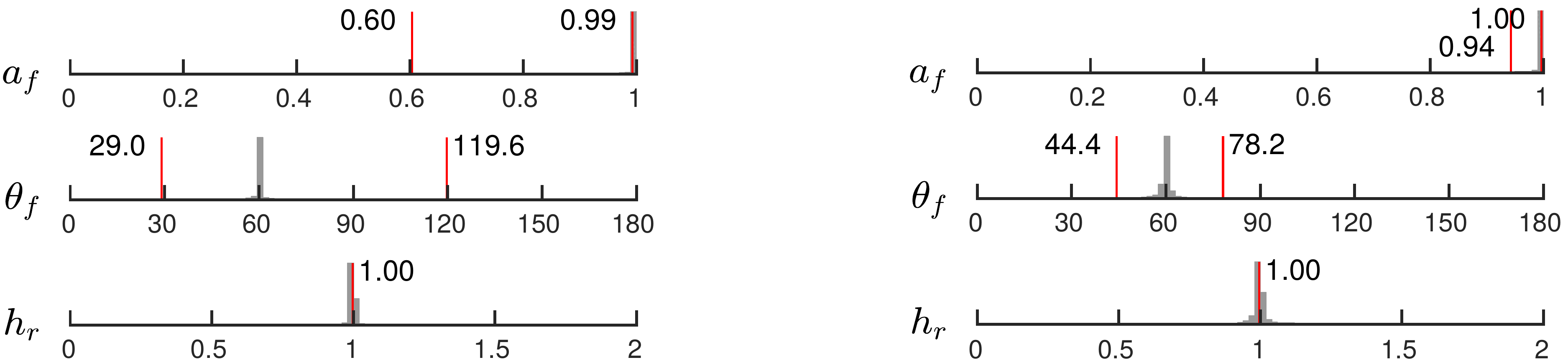}
  
  \caption{Mesh-quality metrics for the regionally-refined grid, before (left) and after (right) the application of mesh optimisation. Histograms of area-length ratio $a_{f}$, enclosed-angle $\theta_{f}$ and relative-length ratio $h_{r}$ are illustrated, with min., max.~and mean values annotated.}
  \label{figure_regional_cost}
\end{figure*}

A set of three benchmark problems were considered. The UNIFORM-SPHERE test-case is a fixed resolution meshing problem on the sphere, suitable for uniformly resolved atmospheric and/or oceanic simulations. The REGIONAL-ATLANTIC test-case explores simple, regional-refinement for ocean-modelling, embedding a high-resolution, eddy-permitting representation of the North Atlantic ocean basin in a coarse global grid. The SOUTHERN-OCEAN test-case is designed to test the multi-resolution capabilities of the algorithm, defining a complex non-uniform grid-spacing function for the Southern Ocean and Antarctic regions. The mesh-spacing constraints for this problem were designed to incorporate a combination of topographic- and regional-refinement. The Voronoi/Delaunay grids for these test-cases are shown in Figures~\ref{figure_const_grid}, \ref{figure_regional_grid}, \ref{figure_topo_grid_full} and \ref{figure_topo_grid_zoom}, with the associated grid-quality statistics presented in Figures~\ref{figure_const_cost}, \ref{figure_regional_cost} and \ref{figure_topo_cost}. The underlying grid-spacing definitions used for the REGIONAL-ATLANTIC and SOUTHREN-OCEAN problems are shown in Figure~\ref{figure_grid_spac}. 

In all test cases, limiting radius-edge ratios were specified, with $\bar{\rho}_{f} = 1.05$. This constraint ensures that the minimum angles in the Delaunay triangulations are bounded, with $\theta_{\text{min}}\geq 28.4^\circ$. For all test problems, detailed statistics on element quality are presented, including histograms of element \textit{area-length ratios} $a_{f}$, \textit{element-angles} $\theta_{f}$, and \textit{relative-edge-lengths} $h_{r}$. These quality metrics are described in detail in Section~\ref{section_grid_quality}. Histograms highlight the minimum, maximum and mean values of the relevant distributions as appropriate.

\subsection{Uniform global grid}

\begin{figure*}
  \centering
  \includegraphics[width=.975\textwidth]{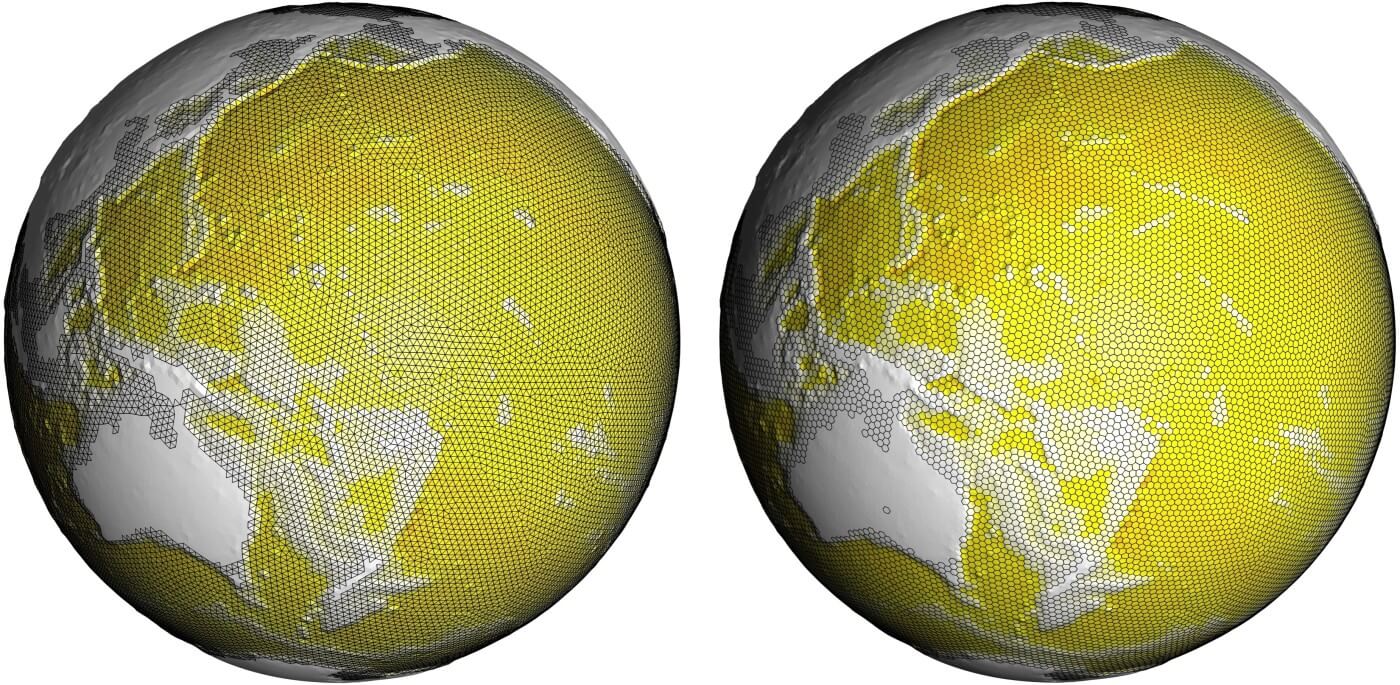}
  
  \caption{A uniform resolution global grid, showing (left) the underlying spheroidal Delaunay triangulation, and (right) the associated staggered Voronoi dual. $150\,\mathrm{km}$ grid-spacing was specified globally. Topography is drawn using an exaggerated scale, with elevation from the reference geoid amplified by a factor of 10.}
  \label{figure_const_grid}
\end{figure*}

\begin{figure*}
  \centering
  
  \includegraphics[width=.865\textwidth]{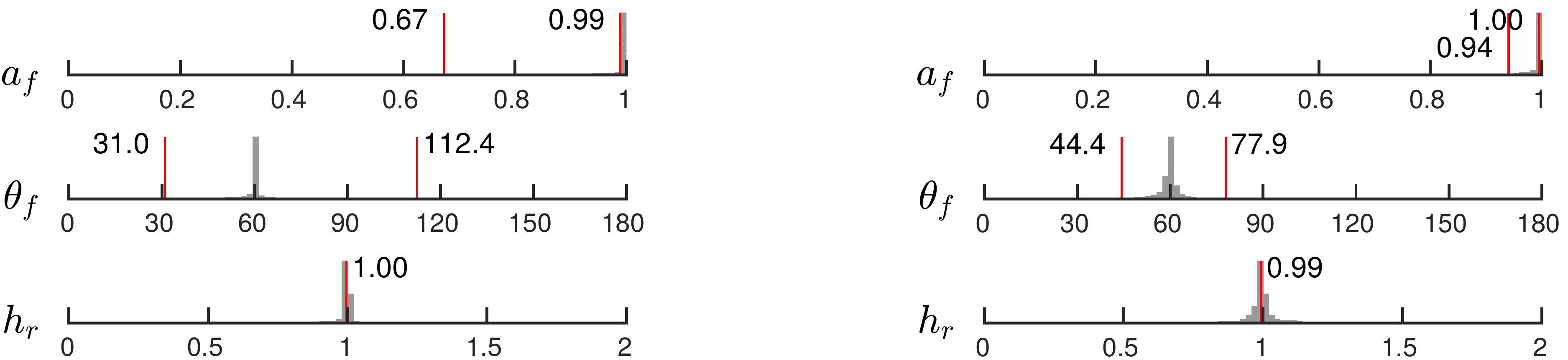}
  
  \caption{Mesh-quality metrics for the uniform resolution grid, before (left) and after (right) the application of mesh optimisation. Histograms of area-length ratio $a_{f}$, enclosed-angle $\theta_{f}$ and relative-length ratio $h_{r}$ are illustrated, with min., max.~and mean values annotated.}
  \label{figure_const_cost}
\end{figure*}

The performance of the JIGSAW-GEO algorithm was first assessed using the UNIFORM-SPHERE test-case, seeking to build a uniformly resolved global grid for general circulation modelling. Spatially uniform mesh-size constraints were enforced, setting $\bar{h}(\mathbf{x}) = 150\,\mathrm{km}$ over the full sphere. The resulting grid is shown in Figure~\ref{figure_const_grid} and contains 83,072 Delaunay triangles and 41,538 Voronoi cells. Grid-quality metrics are presented in Figure~\ref{figure_const_cost}, showing distributions before and after the application of the grid-optimisation procedure.

Visual inspection of these grids show that very high levels of geometric quality are achieved, with both the Delaunay triangulation and Voronoi tessellation consisting of highly regular elements, free of obvious distortion and/or areas of over- or under-refinement. Focusing on the element shape-quality statistics, it is noted that very high levels of mesh regularity are achieved, with the vast majority of element area-length scores tightly clustered about $a_{f}=1$. Similarly, the distribution of element angles shows strong convergence around $\theta_{f}=60^\circ$, revealing most triangles to be near equilateral. Finally, analysis of the relative-length distribution shows that edge-lengths follow the imposed mesh-spacing constraints closely, with very tight clustering about $h_{r}=1$.

The effect of the grid-optimisation procedure can be assessed by comparing the mesh-quality statistics presented in Figure~\ref{figure_const_cost}. The application of mesh-optimisation is seen to be most pronounced at the \textit{tails} of the distributions, showing that, as expected, the hill-climbing type procedure is effective at improving the worst elements in the grid. Specifically, the minimum area-length metric is improved from $a_{f}=0.67$ to $a_{f}=0.94$, and the distribution of element-wise angles is narrowed from $31^\circ \leq \theta_{f} \leq 112^\circ$ to $44^\circ \leq \theta_{f} \leq 78^\circ$. A slight broadening of the mean parts of the distributions is also evident, showing that in some cases, higher-quality elements are slightly compromised to facilitate improvements to their lower-quality neighbours. This behaviour is consistent with the \textit{worst-first} philosophy employed in this study.

Beyond improvements to grid-quality statistics, the impact of mesh-optimisation can be further understood by considering the so-called \textit{well-centredness} of the resulting staggered Voronoi/Delaunay grid. Well-centred triangulations are those for which all element circumcentres are located within their parent triangles, ensuring that the associated Voronoi cells are \textit{nicely-staggered} with respect to the underlying triangulation as a result. Such a constraint is equivalent to requiring that all Delaunay triangles are \textit{acute}, with $\theta_{f} \leq 90^\circ$. Further details regarding these constraints are presented in Section~\ref{section_mpas_grids}. 

Well-centred grids are highly desirable from a numerical perspective, allowing, for instance, the unstructured C-grid scheme employed in the MPAS framework to achieve optimal rates of convergence. Specifically, when a grid is well-centred, it is guaranteed that associated edges in the staggered Voronoi and Delaunay cells intersect, ensuring that evaluation of the cell-transport and circulation terms can be accurately computed using compact numerical stencils. In the case of \textit{perfectly-centred} grids, such intersections occur at edge-midpoints --- allowing a numerical scheme based on local linear interpolants to achieve second-order accuracy. 

The construction of well-centred grids is known to be a difficult problem, and the development of algorithms for their generation is an ongoing area of research \citep{vanderzee2008well,vanderzee2010well}. For the uniform resolution case studied here, it is clear that the JIGSAW-GEO algorithm is successful in generating such a grid, with all angles bounded below $77.9^\circ$.

\subsection{Regionally-refined North Atlantic grid}

\begin{figure*}
  \centering
  \includegraphics[width=.9675\textwidth]{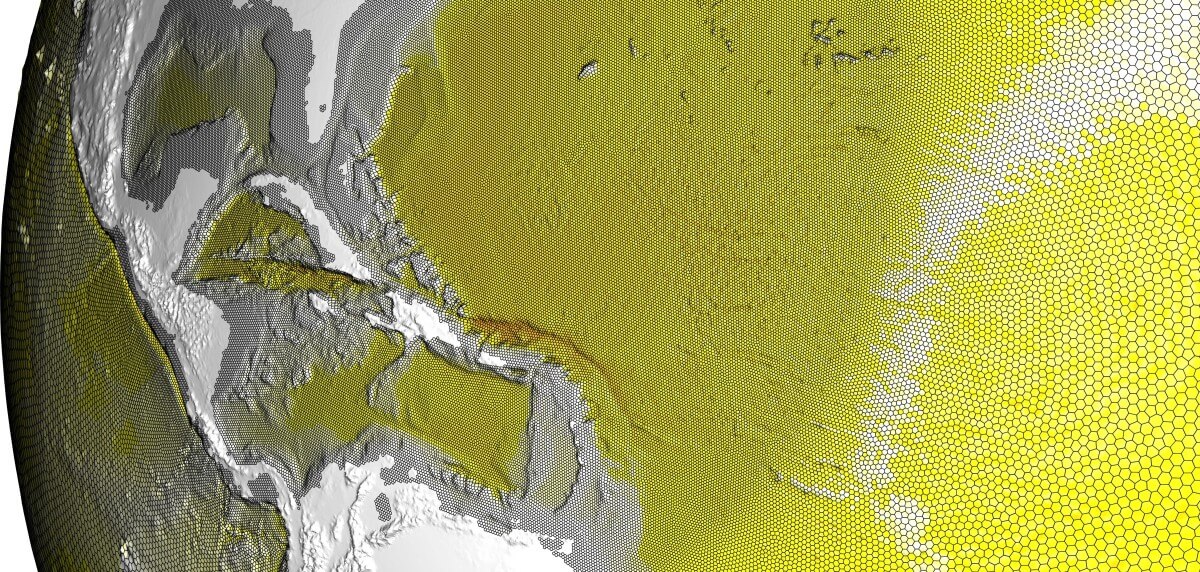}
  
  \caption{Additional detail of the regionally-refined North Atlantic grid shown in Figure~\ref{figure_regional_grid}.}
  \label{figure_regional_grid_zoom}
\end{figure*}

The multi-resolution capabilities of the JIGSAW-GEO algorithm were investigated in the REGIONAL-ATLANTIC test-case, seeking to build a regionally-resolved grid for high-resolution modelling of the North Atlantic ocean basin. Non-uniform mesh-size constraints were enforced, setting $\bar{h}(\mathbf{x}) = 150\,\mathrm{km}$ globally, with $15\,\mathrm{km}$ \textit{eddy-permitting} mesh-spacing specified over the North Atlantic region. The resulting grid is shown in Figure~\ref{figure_regional_grid} and contains 358,064 Delaunay triangles and 179,081 Voronoi cells. Grid-quality metrics are presented in Figure~\ref{figure_regional_cost}, showing distributions before and after the application of grid-optimisation.

Consistent with the results presented previously, a very high-quality Voronoi/Delaunay grid was generated for the REGIONAL-ATLANTIC problem, with each grid-quality metric tightly clustered about its optimal value, such that $a_{f}\rightarrow 1$, $\theta_{f}\rightarrow 60^\circ$ and $h_{r}\rightarrow 1$. As per the uniform resolution case, the application of mesh-optimisation led to significant improvements in worst-case grid quality, with the minimum area-length metric improved from $a_{f}=0.60$ to $a_{f}=0.94$, and the distribution of element angles narrowed from $29^\circ \leq \theta_{f} \leq 120^\circ$ to $44^\circ \leq \theta_{f} \leq 78^\circ$. The resulting optimised grid is also clearly \textit{well-centred}, with all angles in the Delaunay trianglulation less than $78.2^\circ$. Overall, grid-quality can be seen to achieve essentially the same levels of optimality as the uniform resolution test-case, showing that the JIGSAW-GEO algorithm can be used to generate very high-quality spatially-adaptive grids without obvious degradation in mesh-quality.

\subsection{Multi-resolution Southern Ocean grid}

\begin{figure*}
  \centering
  \includegraphics[width=.9675\textwidth]{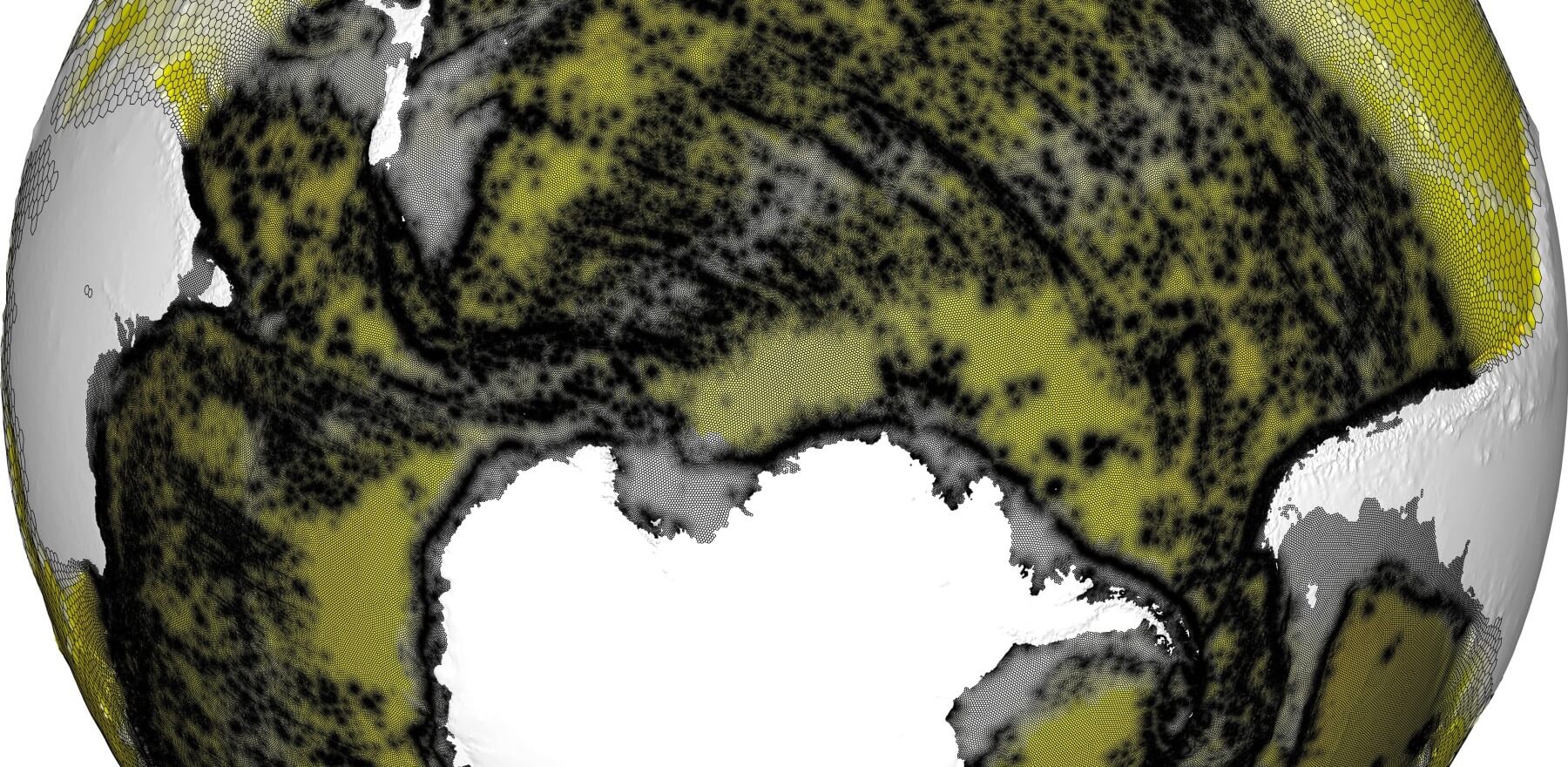}
  
  \caption{A multi-scale grid for the Southern Ocean. Coarse resolution is $150\,\mathrm{km}$, with $15\,\mathrm{km}$ eddy-permitting grid-spacing specified south of $32.5^\circ\,\mathrm{S}$. Local topographic adaptation is also utilised, with resolution increased in areas of large bathymetric gradient. Minimum grid-spacing is $4\,\mathrm{km}$.  Topography is drawn using an exaggerated scale, with elevation from the reference geoid amplified by a factor of 10.}
  \label{figure_topo_grid_full}
\end{figure*}

\begin{figure*}
  \centering
 
  \includegraphics[width=.865\textwidth]{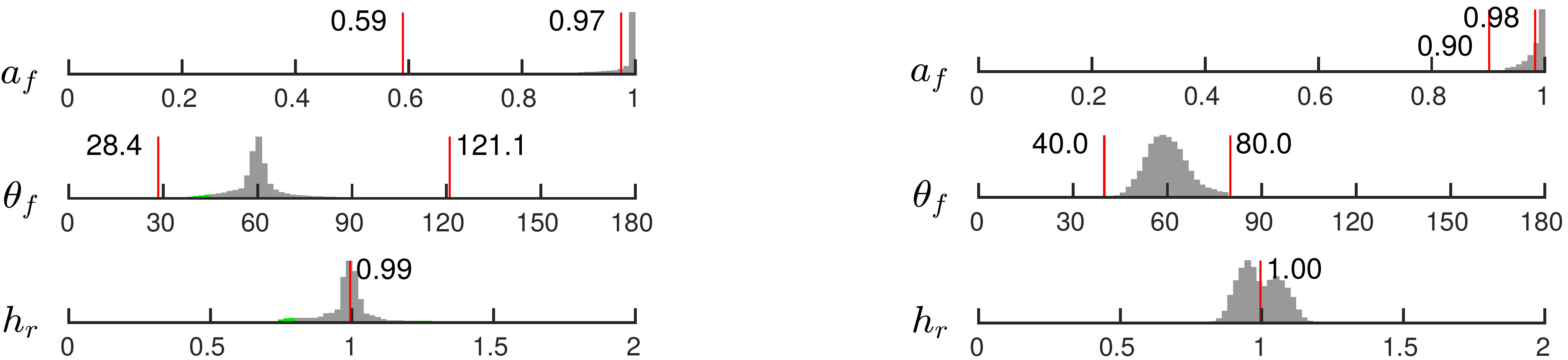}
  
  \caption{Mesh-quality metrics associated with the multi-resolution grid, before (left) and after (right) the application of mesh optimisation. Histograms of area-length ratio $a_{f}$, enclosed-angle $\theta_{f}$ and relative-length ratio $h_{r}$ are illustrated, with min., max.~and mean values annotated.}
  \label{figure_topo_cost}
\end{figure*}

\begin{figure*}
  \centering
 
  \includegraphics[width=1.\textwidth]{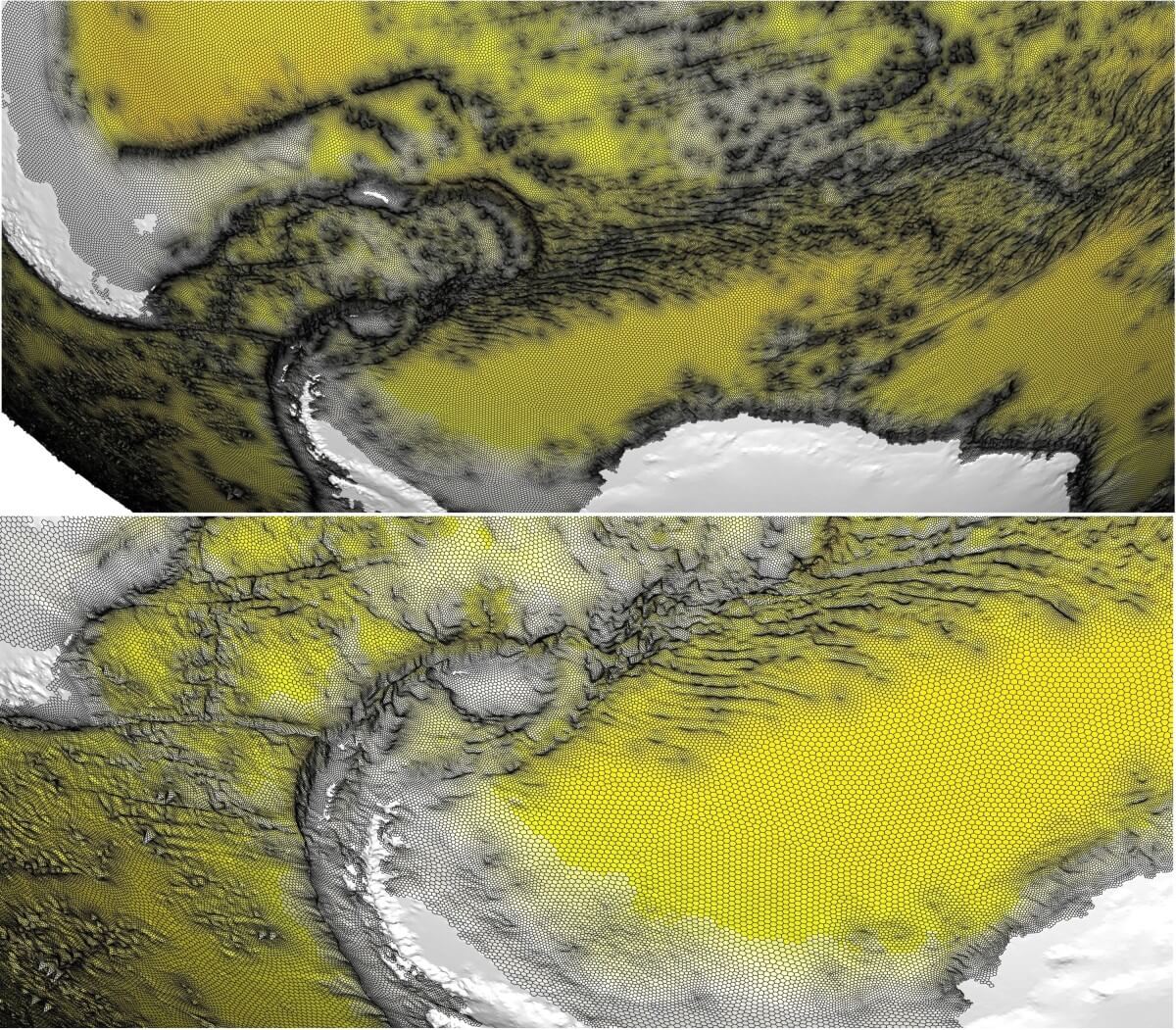}
  
  \caption{Additional detail of the multi-resolution Southern Ocean grid shown in Figure~\ref{figure_topo_grid_full}.}
  \label{figure_topo_grid_zoom}
\end{figure*}

The JIGSAW-GEO algorithm was then used to mesh the challenging SOUTHERN-OCEAN test-case, allowing its performance for large-scale problems involving rapidly-varying mesh-spacing constraints to be analysed in detail. This test-case seeks to build a multi-resolution grid for regionally-refined ocean studies, with a particular focus on resolution of the Antarctic Circumpolar Current (ACC), and adjacent Antarctic processes. Composite mesh-spacing constraints were enforced, consisting of a coarse global background resolution of $150\,\mathrm{km}$, with an \textit{eddy-permitting} $15\,\mathrm{km}$ grid-spacing specified south of $32.5^\circ\mathrm{S}$. Additional topographic adaptation was also utilised within the southern annulus, with grid-resolution increased in regions of large bathymetric gradient. A minimum grid-spacing of $4\,\mathrm{km}$ was specified. Topographic gradients were computed using the high-resolution ETOPO1 Global Relief dataset \citep{amante2009etopo1}. The resulting grid is shown in Figure~\ref{figure_topo_grid_full}, with additional detail shown in Figure~\ref{figure_topo_grid_zoom}. The grid contains 3,119,849 Delaunay triangles and 1,559,927 Voronoi cells. Associated grid-quality metrics are presented in Figure~\ref{figure_topo_cost}, showing distributions before and after the application of the grid-optimisation procedure.

Consistent with previous results, visual inspection of Figures~\ref{figure_topo_grid_full} and \ref{figure_topo_grid_zoom} confirm that the JIGSAW-GEO algorithm is capable of generating very high quality multi-resolution grids, containing a majority of near-perfect Delaunay triangles and Voronoi cells. Additionally, it can be seen that grid resolution varies smoothly, even in regions of rapidly-fluctuating mesh-spacing constraints (see Figure~\ref{figure_topo_grid_zoom} for detail). Analysis of the grid-quality metrics shown in Figure~\ref{figure_topo_cost} show that very high levels of mesh regularity are achieved, with element area-length scores tightly clustered about $a_{f}=1$ and element-angles showing strong convergence around $\theta_{f}=60^\circ$. Interestingly, despite the complexity of the imposed mesh-spacing function, the relative-length distribution still shows relatively tight conformance, with a sharp clustering about $h_{r}=1$. Overall, mean grid-quality is slightly reduced compared to the uniform resolution case, illustrated by a slight broadening of the grid-quality distributions. Note that such behaviour is expected in the multi-resolution case, with imperfect triangle geometries required to satisfy non-uniform mesh-spacing constraints.

The effect of the grid-optimisation procedure can be assessed by comparing the mesh-quality statistics presented in Figure~\ref{figure_topo_cost}. As per the uniform resolution case, mesh-optimisation appears to be most aggressive at the \textit{tails} of the distributions, acting to improve the worst elements in the grid. The minimum area-length metric is improved from $a_{f}=0.59$ to $a_{f}=0.90$, and the distribution of element angles is narrowed from $28^\circ \leq \theta_{f} \leq 121^\circ$ to $40^\circ \leq \theta_{f} \leq 80^\circ$. Consistent with previous results, a moderate broadening of the mean components of the distributions can be observed, especially in the enclosed-angle and relative-length metrics. As per previous discussions, this behaviour shows that some high-quality elements are slightly compromised to improve their lower quality neighbours.

The optimised grid is also clearly \textit{well-centred}, with all angles in the Delaunay triangulation bounded below $80^\circ$. This result shows only marginal degradation compared to the simpler examples presented previously --- demonstrating the effectiveness of the optimisation strategies described here. These results show that very high-quality, well-centred grids can be generated for complex multi-resolution cases, even when the grid-spacing function incorporates strong local fluctuation. Nonetheless, the construction of well-centred grids remains a challenging task, and it is noted that there may exist test-cases that defeat the current strategy. As such, the pursuit of alternative optimisation techniques, designed to target grid well-centredness directly, is an interesting avenue for future research.

\subsection{Computational performance}

In addition to the generation of very high-quality grids, the new JIGSAW-GEO algorithm also imposes a relatively moderate computational burden, producing large-scale, multi-resolution grids in a matter of minutes using standard desktop-based computing infrastructure. Specifically, grid-generation for the UNIFORM-SPHERE, REGIONAL-ATLANTIC and SOUTHERN-OCEAN test-cases required approximately 12 seconds, $1\tfrac{1}{2}$ minutes and 10 minutes of computation time, respectively, running on a single core of an Intel i7 processor. In all cases, grid-optimisation was found to be approximately four times as expensive as the initial Frontal-Delaunay refinement. Compared to the existing iterative MPI-SCVT algorithm \citep{jacobsen2013parallel}, commonly used to generate grids for the MPAS framework, these results represent a significant increase in productivity, with the MPI-SCVT algorithm often requiring days, or even weeks of distributed computing time. 

Additionally, practical experience with the MPI-SCVT algorithm has shown that it cannot always be relied upon to generate an appropriate grid, irrespective of the amount of computational time allowed for convergence to be reached. While always generating \textit{locally-orthogonal} and \textit{centroidal} tessellations with very high mean grid-qualities, the MPI-SCVT algorithm does not provide bounds on the worst-case metrics. In practice, multi-resolution grids generated using the MPI-SCVT algorithm are often observed to contain a minority of obtuse triangles that violate the \textit{well-centred} constraint. As per the discussions presented in Section~\ref{section_mpas_grids}, such grids are inappropriate for use in an unstructured C-grid model such as MPAS. Currently, grid-generation for such models often requires a degree of user-driven \textit{trial-and-error} as a result, making grid-generation a somewhat arduous task for model-users. Initial experiments conducted using the JIGSAW-GEO algorithm have shown it to be a useful alternative, reliably generating valid well-centred multi-resolution grids for a wide range of user-defined constraints and configuration settings. Initial evaluation of JIGSAW-GEO for use with the MPAS framework is currently underway.

\section{Conclusions \& Future Work}
\label{section_conclusion}

A new algorithm for the generation of multi-resolution staggered unstructured grids for large-scale general circulation modelling on the sphere has been described. Using a combination of Frontal-Delaunay refinement and hill-climbing type optimisation techniques, it has been shown that very high-quality \textit{locally-orthogonal}, \textit{centroidal} and \textit{well-centred} spheroidal grids appropriate for unstructured C-grid type general circulation models can be generated. The performance of this new approach has been verified using a number of multi-scale global benchmarks, including difficult problems incorporating highly non-uniform mesh-spacing constraints.

This new algorithm is available as part of the JIGSAW meshing package, providing a simple and easy-to-use tool for the oceanic and atmospheric modelling communities. A number of benchmark problems have been analysed, examining the performance of the new approach. The Frontal-Delaunay refinement algorithm has been shown to generate \textit{guaranteed-quality} spheroidal Delaunay triangulations --- satisfying worst-case bounds on element angles and exhibiting smooth grading characteristics. This algorithm has been shown to produce very high-quality multi-resolution triangulations, with a majority of elements exhibiting strong conformance to element-shape and grid-spacing constraints.

The use of a coupled geometrical and topological hill-climbing optimisation procedure was shown to further improve grid-quality, especially for the lowest quality elements in each mesh. It was demonstrated that these optimisation techniques allow grid-quality to be improved to the extent that fully \textit{well-centred} mesh configurations can be achieved, with angles in the Delaunay triangulation bounded below $90^\circ$. For the three global test-cases presented here, angles were bounded above $\theta_{f} \geq 40^\circ$ and below $\theta_{f} \leq 80^\circ$. 

The construction of meshes appropriate for a range of contemporary unstructured C-grid type general circulation models was also discussed in detail, with a focus on the generation of multi-resolution grids for the MPAS framework. The availability of this new algorithm is expected to significantly reduce the grid-generation burden for MPAS model-users. 

Future work will focus on a generalisation of the algorithm and improvements to its efficiency, including: (i) support for coastal constraints, (ii) improvements to computational performance through parallelism, and (iii) further enhancements to the mesh optimisation procedures, with a focus on improving the \textit{well-centredness} of the resulting staggered grids. The investigation of \textit{solution-adaptive} multi-scale representations, in which grid-resolution is adapted to spatial variability in model state \citep{JAME20286}, is also an obvious direction for future investigation.

\begin{acknowledgements} 
This work was carried out at the NASA Goddard Institute for Space Studies, the Massachusetts Institute of Technology, and the University of Sydney with the support of a NASA--MIT cooperative agreement and an Australian Postgraduate Award. The author wishes to thank Todd Ringler, Luke Van Roekel, Mark Petersen, Matthew Hoffman and Phillip Wolfram for their assistance on grid-generation for the MPAS-O environment. John Marshall provided feedback on an earlier version of the manuscript. The author also wishes to thank the anonymous reviewers for their helpful comments and feedback. 
\end{acknowledgements}

\section{Code availability}

The JIGSAW-GEO grid-generator used in this study is available as a Zenodo archive: \url{https:
//doi.org/10.5281/zenodo.556602}. The JIGSAW-GEO framework is under active development and the latest version can be accessed here: \url{https://github.com/dengwirda/jigsaw-geo-matlab}.


\appendix
\section{Spheroidal Predicates}
\label{appendix_predicates}

Computation of the \textit{restricted} Delaunay surface tessellation $\operatorname{Del}|_{\Sigma}(X)$ requires the evaluation of a single \textit{predicate}. Given a spheroidal surface $\Sigma$, the task is to compute intersections between edges in the Voronoi tessellation $\operatorname{Vor}(X)$ and the surface $\Sigma$.

\subsection{Restricted surface triangles}

Restricted surface triangles $f_{i}\in\operatorname{Del}|_{\Sigma}(X)$ are defined as those associated with an \textit{intersecting} Voronoi edge $v_{e}\in\operatorname{Vor}(X)$, where $v_{e}\cap\Sigma\neq\emptyset$. These triangles provide a good piecewise linear approximation to the surface $\Sigma$. For a given triangle $f_{i}$, the associated Voronoi edge $v_{e}$ is defined as the line-segment joining the two \textit{circumcentres} $\mathbf{c}_{i}$ and $\mathbf{c}_{j}$ associated with the pair of tetrahedrons that share the face $f_{i}$. The task then is to find intersections between the line-segments $v_{e}$ and the surface $\Sigma$. Let $\mathbf{p}$ be a point on a given Voronoi edge-segment $v_{e}$
\begin{gather}
\label{equation_voronoi_edge}
\qquad \mathbf{p}  = \bar{\mathbf{c}} + t \Delta \,, \quad {-1} \leq t \leq {+1} \,,
\quad \text{where}
\\[1ex]
\qquad \bar{\mathbf{c}} = \tfrac{1}{2}\left(\mathbf{c}_{i}+\mathbf{c_{j}}\right)
\quad\text{and}
\quad \Delta = \tfrac{1}{2}\left(\mathbf{c}_{j}-\mathbf{c_{i}}\right) \,.
\end{gather}
Substituting (\ref{equation_voronoi_edge}) into the equation of the spheroidal surface $\Sigma$ leads to a quadratic expression for the parameter $t$
\begin{gather}
\qquad \sum_{i=1}^{3} \, \left(\frac{ \bar{\mathbf{c}}_{i} + t\Delta_{i} }{r_{i}}\right)^{2} = 1 \,,
\\[1ex]
\qquad \sum_{i=1}^{3} \, \frac{ \bar{\mathbf{c}}_{i}^{2} + 2 t \bar{\mathbf{c}}_{i}\Delta_{i} + t^{2}\Delta_{i}^{2} }{r_{i}^{2}} = 1 \,,
\\[1ex]
\qquad \sum_{i=1}^{3} \, \left(\frac{\Delta_{i}^{2}}{r_{i}^{2}}\right) t^{2} +
\left(\frac{2\bar{\mathbf{c}}_{i}\Delta_{i}}{r_{i}^{2}}\right) t + \left(\frac{\bar{\mathbf{c}}_{i}^{2}}{r_{i}^{2}} - 1\right) = 0 \,.
\end{gather}
Any real solutions ${-1} \leq t_{\Sigma} \leq {+1}$ correspond to non-trivial intersections $v_{e}\cap\Sigma\neq\emptyset$. The corresponding point of intersection $\mathbf{p}_{\Sigma}$ can be found by substituting $t_{\Sigma}$ into (\ref{equation_voronoi_edge}).

\end{document}